\documentclass{aa}  

\usepackage{graphicx}
\usepackage{color}

\usepackage{txfonts}
\begin{document} 

   \title{The VMC Survey - XXXII. Pre-main sequence populations in the Large Magellanic Cloud}

   \author{Viktor Zivkov \inst{1,2},
          Joana M. Oliveira \inst{1},
          Monika G. Petr-Gotzens\inst{2},
          Maria-Rosa L. Cioni\inst{3},
          Stefano Rubele\inst{4,5},
          Jacco Th. van Loon\inst{1},
          Kenji Bekki\inst{6},
          Felice Cusano\inst{7},
          Richard de Grijs\inst{8,9,10},
          Valentin D. Ivanov\inst{2},
          Marcella Marconi\inst{11},
          Florian Niederhofer\inst{3},
          Vincenzo Ripepi\inst{11},
          Ning-Chen Sun\inst{12}
          }

   \institute{Lennard-Jones Laboratories, School of Chemical and Physical Sciences,
   		Keele University, ST5 5BG, UK
         \and
             European Southern Observatory, Karl-Schwarzschild-Str. 2,
             85748 Garching bei M\"unchen, Germany
          \and
          Leibniz-Institut f\"ur Astrophysik Potsdam, An der Sternwarte 16, D-14482 Potsdam, Germany
          \and
          	Dipartimento di Fisica e Astronomia, Universit\`a di Padova,
         	Vicolo dell’Osservatorio 2, I-35122 Padova, Italy
          \and
          	Osservatorio Astronomico di Padova – INAF,
            Vicolo dell’Osservatorio 5, I-35122 Padova, Italy
          \and 
          	ICRAR, M468, University of Western Australia, 35 Stirling Hwy, 6009 Crawley, 			Western Australia, Australia
           \and 
          INAF-Osservatorio di Astrofisica e Scienza dello Spazio di Bologna, Via
P. Gobetti 93/3, I-40129 Bologna, Italy
          \and 
          	Department of Physics and Astronomy, Macquarie University, Balaclava Road, Sydney, NSW 			2109, Australia
          \and 
          Research Centre for Astronomy, Astrophysics and Astrophotonics, Macquarie University, Balaclava Road, Sydney, NSW 2109, Australia
          \and 
          International Space Science Institute--Beijing, 1 Nanertiao, Zhongguancun, Hai Dian District, Beijing 100190, China
          \and 
          	INAF-Osservatorio Astronomico di Capodimonte, via Moiariello 16, 80131, 				Naples, Italy
          \and 
          	Kavli Institute for Astronomy \& Astrophysics and Department of Astronomy,
          	Peking University, Yi He Yuan Lu 5, Hai Dian District, Beijing 100871, China}

   \date{}

 
  \abstract
   {Detailed studies of intermediate/low mass pre-main sequence (PMS) stars outside the Galaxy have so far been conducted only for small targeted regions harbouring known star formation complexes. The VISTA Survey of the Magellanic Clouds (VMC)
provides an opportunity to study PMS populations down to solar masses on a galaxy-wide scale.}
   {Our goal is to use near-infrared data from the VMC survey to identify and characterise PMS populations down to $\sim 1\,\mathrm{M_{\sun}}$ across the Magellanic Clouds. We present our colour$-$magnitude diagram method, and apply it to a $\sim 1.5\,\textrm{deg}^2$ pilot field located in the Large Magellanic Cloud.}
   {The pilot field is divided into equally-sized grid elements. We compare the stellar population in every element with the population in nearby control fields by creating $K_s/(Y-K_s)$ Hess diagrams; the observed density excesses over the local field population are used to classify the stellar populations.}
   {Our analysis recovers all known star formation complexes in this pilot field (N\,44, N\,51, N\,148 and N\,138) and for the first time reveals their true spatial extent. In total, around 2260 PMS candidates with ages $\lesssim 10$\,Myr are found in the pilot field. PMS structures, identified as areas with a significant density excess of PMS candidates, display a power-law distribution of the number of members with a slope of $-0.86\pm0.12$. We find a clustering of the young stellar populations along ridges and filaments where dust emission in the far-infrared (FIR) ($70\,\mathrm{\mu m}$ -- $500\,\mathrm{\mu m}$) is bright. Regions with young populations lacking massive stars show a lesser degree of clustering and are usually located in the outskirts of the star formation complexes. At short FIR wavelengths ($70\,\mathrm{\mu m}$, $100\,\mathrm{\mu m}$) we report a strong dust emission increase in regions hosting young massive stars, which is less pronounced in regions populated only by less massive ($\lesssim 4\,\mathrm{M_{\sun}}$) PMS stars.}
   {}

   \titlerunning{PMS populations in the LMC}
   \authorrunning{Zivkov et al.}
          
   \maketitle
%

\section{Introduction}
\label{sec:intro}

The Magellanic Clouds (MCs) are nearby, interacting, gas-rich galaxies with metallicities lower than typically encountered in the Milky Way \citep{Stanimirovic2004,Besla2012}. With moderate distances of $50\pm2$\,kpc for the Large Magellanic Cloud (LMC; \citealt{Grijs2014}) and $61.9\pm0.6$\,kpc for the Small Magellanic Cloud (SMC; \citealt{Grijs2015}), they provide a great opportunity to study resolved star formation down to the scales of individual young stellar objects under different environmental conditions than those found in the Galaxy.
The LMC has an apparent size on the sky of approximately $5.4\degr\times 4.6\degr$ \citep{Cook2014} and is seen almost face-on \citep{vanderMarel2001}. Its depth is $4.0\pm1.4$\,kpc and $3.44\pm1.16$\,kpc in the bar region and the disc, respectively \citep{Subramanian2009}, leading to relatively small distance modulus variations among its stellar members. The mean metallicity of the LMC is approximately half of the solar metallicity \citep{Russell1992}, which puts it close to the mean metallicity of the interstellar medium during the time of peak star formation in the Universe \citep{Pei1999}. 

The first studies of young stellar populations within the MCs were published over half a century ago \citep{Westerlund1961, Bok1964}. \cite{Lucke1970} created a catalogue of 122 OB associations in the LMC, based on observations at optical wavelengths down to $m_V \approx 16$\,mag, which corresponds at the LMC distance to an $\sim11\,\mathrm{M_{\sun}}$ main sequence star of age 10\,Myr \citep{Bressan2012}. The advent of sensitive CCD detector arrays improved the detection limit so that more detailed studies of the stellar content of associations became possible \citep[e.g.][]{Massey1989b, Massey1989a}.
A huge leap in sensitivity and resolution was provided by the \textit{Hubble Space Telescope} (HST). Optical imaging studies of young clusters and OB associations in both galaxies have found evidence of extensive pre-main sequence (PMS) populations well below the solar mass regime \citep[e.g.][]{Gilmozzi1994, Panagia2000, Gouliermis2006, Gouliermis2006b, Gouliermis2007, Gouliermis2011}. Analysis of the PMS populations in the 30\,Doradus region, revealed by the Hubble Tarantula Treasury Project \citep[HTTP;][]{Sabbi2013}, helped to reconstruct the star formation history of the complex \citep{Cignoni2015}. These authors reported that low-mass stars can form in starburst regions as well as in low density environments. Overall the observed  initial mass function (IMF) in the LMC is consistent with that of the Galaxy \citep[e.g.][]{Dario2009, Liu2009a, Liu2009b}. Narrow-band photometry with an $\mathrm{H\alpha}$-filter was used to identify PMS objects actively undergoing mass accretion \citep{DeMarchi2010, DeMarchi2013, Spezzi2012}. These authors find that PMS stars in the LMC have higher mass accretion rates than stars of similar mass in the Galaxy, a possible effect of metallicity. All these studies targeted individual associations and/or star forming complexes, uncovering the young stellar populations down to masses as low as $\sim 0.3\,\mathrm{M_\sun}$

In contrast, large-scale photometric surveys of the LMC and/or SMC give a galaxy-wide overview of the stellar populations. They allow to investigate the large scale distribution of star forming complexes and their relationship with the underlying gas and dust distribution.
Optical imaging was performed by the Magellanic Cloud Photometric Survey  \citep[MCPS;][]{Zaritsky2002,Zaritsky2004} and near-infrared (NIR) images were provided by the 2\,Micron All Sky Survey \citep[2MASS;][]{Skrutskie2006}. Mid- and far-infrared imaging was performed by the Spitzer Space Telescope \citep[{\it Spitzer};][]{Werner2004} and the Herschel Space Observatory \citep[{\it Herschel};][]{Pilbratt2010} as part of two legacy surveys (SAGE and SAGE-SMC: \citealt{Meixner2006, Gordon2011}; HERITAGE: \citealt{Meixner2013}). These surveys provided a comprehensive overview of the high-mass young stellar content and their spatial distribution. However they lack the depth and resolution to study the intermediate/low mass young stellar population.

The VISTA Survey of the Magellanic Clouds \citep[VMC;][]{Cioni2011} provides a significant improvement in depth and resolution. Its data are being used, amongst other works, to characterise the stellar content of the MCs. \cite{Piatti2014} analysed the colour$-$magnitude diagrams (CMDs) of known clusters in the LMC, while \cite{Piatti2016} used stellar over-densities to detect new stellar clusters in the SMC with ages between $\log{(t/\mathrm{yr}) \sim 7.5-9.0}$. \cite{Sun2017b,Sun2017a,Sun2018} used upper main-sequence (UMS) stars to trace large scale structures in major star formation complexes in the LMC and SMC. They found that the size and mass distributions follow a power law, which supports hierarchical star formation governed by turbulence.
Less massive PMS stars, due to their extended PMS phase \citep{Baraffe2015}, provide valuable information about the recent star formation history \citep[see][for an overview]{Gouliermis2012}.
Identifying the PMS populations, including star formation sites composed of intermediate/low mass stars only, can thus reveal the full galaxy-wide extent of recent star formation.

We present an automated method that uses the capabilities of the VMC to detect intermediate/low mass ($1\,\mathrm{M_\sun} \lesssim M_* \lesssim 4\,\mathrm{M_\sun}$) PMS populations. 
The method is based on colour$-$magnitude Hess diagram (Hess CMD) analysis, which includes corrections for reddening and completeness, in order to disentangle young populations from the field. This paper describes the development of this method as well as its application to
a $\sim 1.5\,\mathrm{deg}^2$ region in the LMC. The chosen pilot field contains well-studied OB associations with known PMS populations that are used to calibrate and fine-tune the method.
The paper is organized as follows: Sec.\,\ref{sec:dataselect} gives an overview of the VMC survey and the data used in this work. Section\,\ref{sec:decon} describes how we deal with the contamination from old field stars, while Sec.\,\ref{sec:strategy} explains in detail the strategy devised to identify and categorise young populations. In Sec.\,\ref{sec:testing} we present tests using synthetic and literature clusters which evaluate the method's sensitivity depending on the cluster's age and mass. We show in Sec. \ref{sec:results} first results based on the application of our method to the LMC pilot field and discuss the properties of the identified young low-mass populations. Finally, we present a summary and main conclusions in Sec. \ref{sec:summary}.

\section{VISTA/VMC survey data selection}
\label{sec:dataselect}
The data used in this work are part of the VMC, which is an ESO public survey carried out with the VIRCAM (VISTA InfraRed CAMera) instrument on the 4.1\,m VISTA telescope \citep{Sutherland2015}. The observing strategy of the VMC involves multi-epoch imaging of tiles across the Magellanic System in the $Y$ ($1.02\,\mathrm{\mu m}$), $J$ ($1.25\,\mathrm{\mu m}$), and $K_s$ ($2.15\,\mathrm{\mu m}$) bands. One tile covers almost uniformly an area of $\sim$\,1.5\,$\mathrm{deg^2}$ as a result of combining six offset paw-print images in order to fill the gaps between the 16 VIRCAM detectors. Overall, the survey consists of 110 tiles which cover an area of  $\sim$\,170\,$\mathrm{deg^2}$.  Upon completion every tile will have been observed at 3 epochs in the $Y$ and $J$ bands, and 12 epochs in the $K_s$ band. The exposure time per epoch is 800\,s in $Y$ and $J$, and 750\,s in $K_s$, leading to total exposure times of 2400\,s ($Y$, $J$) and 9000\,s ($K_s$).

\begin{figure}
		\resizebox{\hsize}{!}{\includegraphics{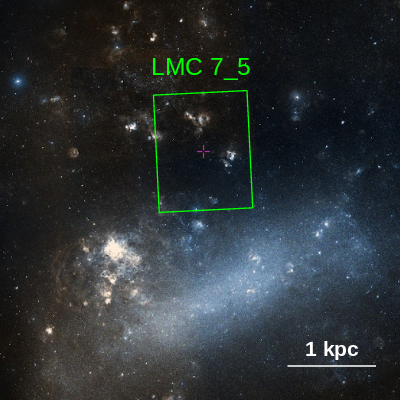}}
		\caption{Digitized Sky Survey (DSS) image showing the location of the pilot field 
(highlighted in green) within the wider LMC environment. North is up and East is to the left.}
		\label{fig:pilot_field}
	\end{figure}

We chose as a pilot field the area covered by the tile LMC\,7\_5. Its central coordinates are $\alpha\,\mathrm{(J2000)} \approx 81\fdg493$ and $\delta\,\mathrm{(J2000)} \approx -67\fdg895$, which is approximately $1\fdg7$ to the Northwest of the Tarantula Nebula and to the North of the LMC bar. The coordinate range covered by this tile is approximately
$80\fdg0 \lesssim \alpha \lesssim 83\fdg0$ and $-68\fdg6 \lesssim \delta \lesssim -67\fdg2.$
Figure \ref{fig:pilot_field} shows the location of tile LMC\,7\_5 within the wider environment of the LMC. This field harbours LHA\,120-N\,44 and  LHA\,120-N\,51 \citep{Henize1956}, two large star forming complexes \citep[e.g.][]{Carlson2012} which are discussed in more detail in Sec.\,\ref{subsec:spa_distribution}. They include massive OB associations like LH\,60 and LH\,63, in which significant populations of intermediate and low mass PMS stars down to $\sim 0.5\,\mathrm{M_\sun}$ have been identified \citep{Gouliermis2011}. Several older clusters with ages between 10\,Myr and 1\,Gyr are also found in this field \citep[e.g.][]{Glatt2010, Popescu2012}.

We use a photometric catalogue obtained by performing point spread function (PSF) photometry on stacked PSF-homogenized images \citep{Rubele2012, Rubele2015}. Previous studies have shown that PSF photometry recovers more sources in crowded regions than aperture photometry \citep[e.g.][]{Tatton2013}. PSF photometry results in deeper and more complete catalogues especially in areas with active or very recent star formation. We tested this by comparing the PSF photometry data with aperture photometry data, reduced and calibrated with the VISTA Data Flow System (VDFS) pipeline \citep{Irwin2004, Gonzales2018} and retrieved from the VISTA Science Archive \citep[VSA\footnotemark; ][]{Cross2012},\footnotetext{http://horus.roe.ac.uk/vsa/} for six circular regions with a radius of $2\arcmin$. Three regions contain young OB associations (LH\,54, LH\,60, LH\,63), while the other three showed no sign of recent star formation activity or stellar over-densities. The PSF photometry detects $\sim 1.4$ times more sources in regions without active star formation, and $\sim 1.6$ times more sources in the regions containing the three young associations. In addition, the PSF catalogues provide estimates of the local completeness for every filter based on artificial star tests \citep{Rubele2012}. These estimates are used to correct for differences in the completeness between stellar populations from different regions (Sec. \ref{subsec:maps}). For the PSF catalogue a completeness of 50\,\% is typically reached at $Y\approx 21.4$\,mag, $J\approx 21.3$\,mag, and $K_s\approx 20.6$\,mag. The $5\sigma$ magnitude limits are $Y\approx 22.3$\,mag, $J\approx 21.9$\,mag, and $K_s\approx 20.9$\,mag \citep[for the nominal aperture photometry magnitude limits, see][]{Cioni2011}. Using the mean reddening in LMC\,7\_5 ($E(Y-K_s)\approx 0.18$\,mag; Tatton et al., in prep.), the PSF $5\sigma$ limits correspond to $\sim0.7\,\mathrm{M_{\sun}}$ and $\sim1.3\,\mathrm{M_{\sun}}$ for ages of 1\,Myr and 10\,Myr, respectively.

\section{Constructing differential Hess diagrams}
\label{sec:decon}

To identify young stellar populations in the pilot field we use the stars' positions in the CMD which are indicative of their masses and ages.
CMD-based methods  are widely used to analyse clusters of all ages \citep[e.g.][]{Dario2009, Rubele2011, Girardi2013, Niederhofer2017}. In general, observations towards clusters or associations are contaminated by the dispersed field population; it is thus necessary to apply a robust decontamination procedure.
In this paper we work with $K_s/(Y-K_s)$ CMDs, since the longer wavelength baseline makes different populations easier to distinguish.
  
 \subsection{CMD density diagrams}
    \label{subsec:hess}
\begin{figure}
		\resizebox{\hsize}{!}{\includegraphics{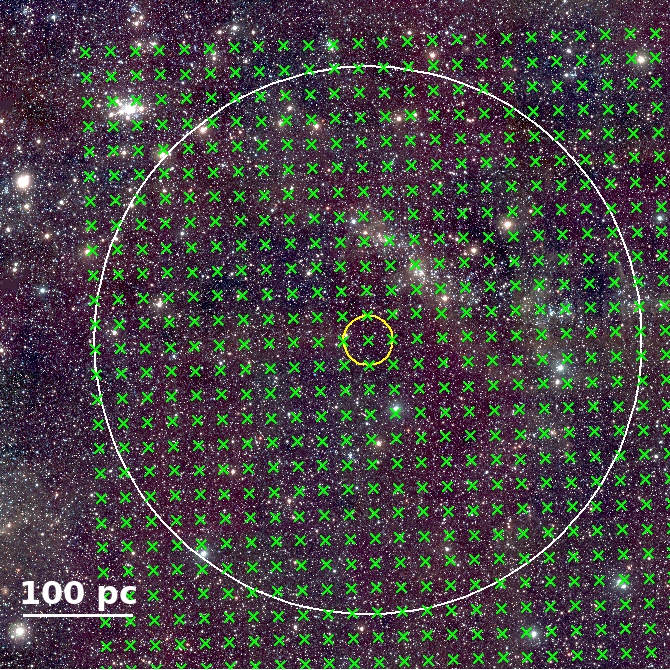}}
		\caption{VISTA RGB composite with $Y$ ($1.02\,\mathrm{\mu m}$) in blue, $J$ ($1.25\,\mathrm{\mu m}$) in green, and $K_s$ ($2.15\,\mathrm{\mu m}$) in red showing the north-eastern corner of the pilot field. Green crosses are placed at the centres of the grid elements ($90\arcsec$ radius in this instance). The small yellow circle highlights an example grid element, while the large white circle shows the area searched to identify suitable control field regions (see text).}
		\label{fig:grid}
	\end{figure}

We start by spatially dividing the pilot field into a grid (see Fig.\,\ref{fig:grid}). Every grid element is circular and the distance between the centres of two neighbouring elements is one grid element radius. The resulting overlap ensures that every location is covered by the grid. It also reduces the chance that a young cluster or association is split up between two or more neighbouring grid elements. The grid radius is a very important parameter and it is the subject of extensive testing in this study (see Sec. \ref{sec:testing}). 
We defined grids with radii of $r = \{90\arcsec,75\arcsec,60\arcsec,50\arcsec,40\arcsec\}$. For each grid element individual $K_s/(Y-K_s)$ CMDs are constructed by extracting the sources' photometry from the PSF catalogue. The CMDs are then smoothed using a Gaussian kernel, resulting in a 2D density map of the stellar distribution in colour$-$magnitude space. The widths of the kernels define the colour$-$magnitude resolution of our density maps; they must be small enough to highlight the distribution of different stellar populations in the CMD, but large enough to be robust against small number statistics. We apply an adaptive kernel width that depends on the number of stars in a grid element ($N_{*, \mathrm{ge}}$) and on the photometric errors as follows:
\begin{equation}
	\label{eq:kernelwidth_one}
	\sigma_{Ks} = 0.2\,\mathrm{mag}\times\sqrt[]{\frac{\langle N_*\rangle}{N_{*, \mathrm{ge}}}}
\end{equation}
\begin{equation}
	\label{eq:kernelwidth_two}
	\sigma_{Y-Ks} = 0.5 \times \sigma_{Ks},
\end{equation}
where $\langle N_*\rangle$ is the median number of stars in the coarsest grid. If the mean photometric error within the kernel, $\Delta\,K_s\,>\,\sigma_{Ks}$ we adopt $\sigma_{Ks}\,=\,\Delta K_s$.
The width values are typically in the ranges $0.1\,\mathrm{mag} < \sigma_{Y-K_s} < 0.2\,\mathrm{mag}$ and $0.2\,\mathrm{mag} < \sigma_{K_s} < 0.4\,\mathrm{mag}$. Kernel width maxima of $\sigma_{Y-K_s}=0.2\,$mag and $\sigma_{K_s}=0.4\,$mag are set to prevent smoothing over too large CMD regions. The smoothing procedure results in a Hess CMD for every grid element. Figure\,\ref{fig:hess_basic} shows representative CMDs (top) and Hess CMDs (bottom) for two grid elements containing mostly old field stars (left) and a star-forming region (right), as shown by the overplotted PARSEC\footnotemark\ isochrones \citep{Bressan2012}. \footnotetext{PAdova and TRieste Stellar Evolution Code \\http://stev.oapd.inaf.it/cgi-bin/cmd\_2.7}

 \subsection{Control field selection}
	\label{subsec:cfields}
\begin{figure}
		\resizebox{\hsize}{!}{\includegraphics{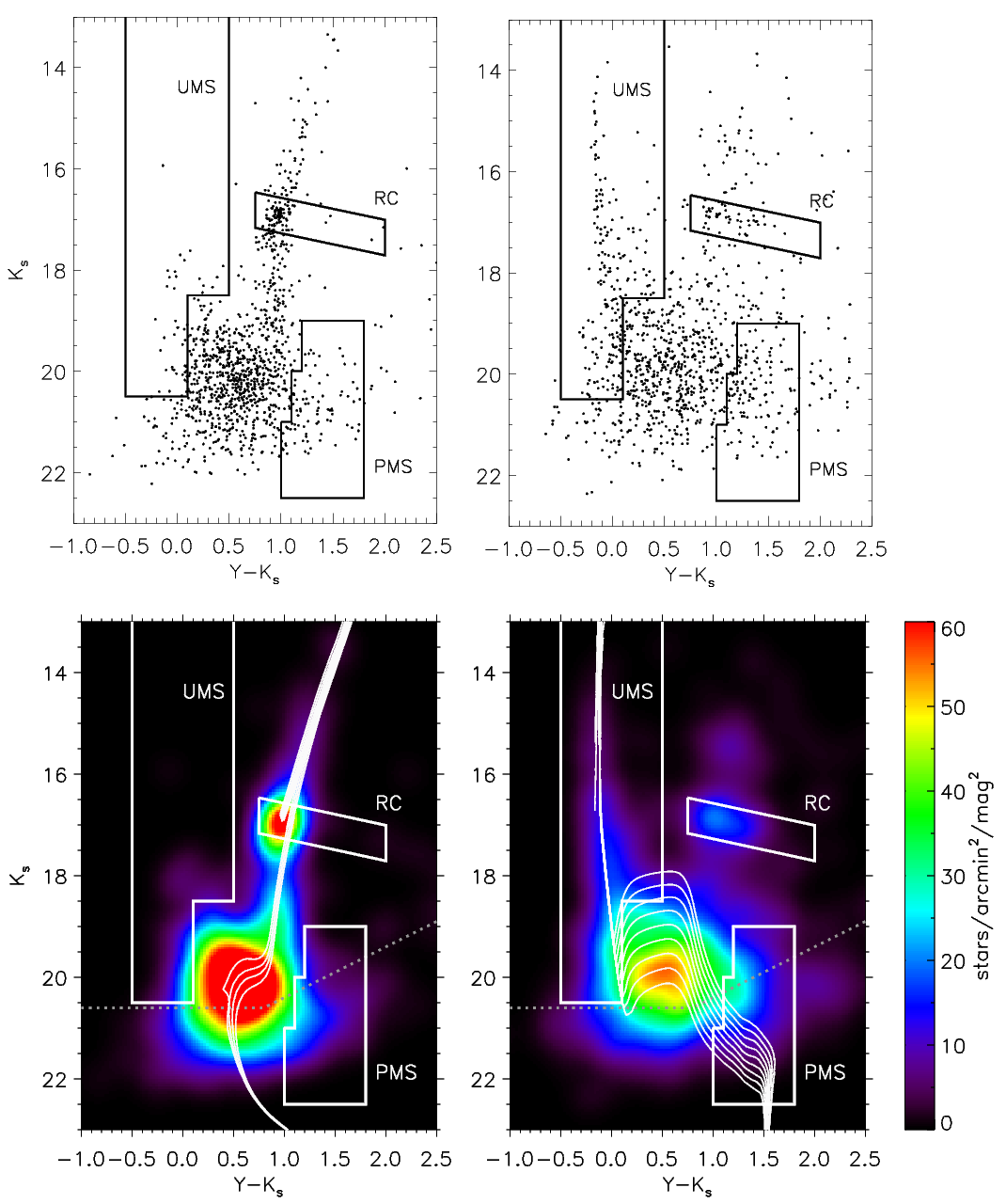}}
		\caption{\textbf{Top:} CMDs of two example grid elements. Boxes indicate the UMS and PMS regions used  in the control field selection, as well as the RC region used to determine the mean extinction (see Sec.\,\ref{subsec:cfields} for details). The slope of the RC selection box is defined by the reddening vector.
          \textbf{Bottom:} Corresponding Hess CMDs.  Thin solid lines are PARSEC isochrones \citep{Bressan2012} for $\log{(t/\mathrm{yr})}\,\in\,[9.5, 9.6, 9.7, 9.8]$ and $Z=0.0033$ (left panel), and $\log{(t/\mathrm{yr})}\,\in\,[6.0, 6.1, 6.2, 6.3, 6.4, 6.5, 6.6, 6.7, 6.8]$ and $Z=0.008$ (right panel). The assumed metallicities are typical for the age ranges \citep{Rubele2012,Tatton2013}. The isochrones are shifted by a distance modulus of 18.49\,mag \citep{Grijs2014}, and an extinction correction derived from the RC analysis (see text) is applied.
        The dotted line marks the typical 50\% completeness level.}
		\label{fig:hess_basic}
	\end{figure}

To decontaminate each grid element from field stars, we use offset control fields which closely resemble the local field population. The total size of the pilot field precludes the use of a single control field for the whole tile, since the typical field population  changes over such large spatial scales. Hence it is necessary to define control fields individually for every grid element. The approach is as follows: we count stars within the UMS and PMS regions of the CMD (shown in Fig.\,\ref{fig:hess_basic}) for all grid elements within a distance of $1000\arcsec$ from the grid element being analysed. The UMS and PMS regions are defined primarily  using PARSEC isochrones. Furthermore, the red limit for the PMS region excludes background galaxies \citep{Kerber2009}, while the blue edge excludes most old field stars. The boundary at bright magnitudes is generous enough to allow for effects like PMS variability and IR excesses, and the lower boundary is imposed by the sensitivity of the survey.
The CMD of a typical LMC field population is expected to have significantly fewer stars within the UMS and PMS regions than that of  an area dominated by young stars. Star counts can thus identify grid elements that are control field candidates. An example control field search area is depicted in Fig.\,\ref{fig:grid}. The number of grid elements within this area varies, depending on the size of the elements; the coarsest grid ($90\arcsec$ radius) contains around 360 grid elements. We create histograms of the star counts for both the UMS and PMS regions and approximate these with Gaussian distributions.

In addition, the mean extinction towards every grid element within the control field search area is determined by using the mean observed colour of red clump (RC) stars. These are evolved stars with well-constrained luminosities and thus the RC is often used for distance  and reddening measurements \citep[e.g.][]{Paczynski1998, Tatton2013}. Although population effects change the absolute magnitude of RC stars \citep{Girardi2001}, any variations on $\ll$\,kpc scales are generally dominated by distance differences and reddening. 
The selection box for the RC is also indicated in Fig.\,\ref{fig:hess_basic}.
Its slope of 0.434 is calculated based on the relative extinctions for VISTA filters, $A_{Y}/A_{V}\approx 0.390$ and $A_{K_s}/A_{V}\approx 0.118$ \citep{Catelan2011}. The $Y-K_s$ colour limits were chosen to keep the contamination by non-RC objects low, while still covering typical LMC reddening values \citep{Rubele2012,Tatton2013}. Assuming an intrinsic RC colour of $(Y-K_s)_0 =0.84$\,mag \citep{Tatton2013}, we probe extinction values up to $A_V \approx 4.26$\,mag. 
The magnitude limits are such as to be insensitive to the small distance variations due to the depth of the LMC along the line of sight. We create a histogram of all mean extinction values within the control field search area, and fit it with a Gaussian distribution. Across the pilot field the mean extinction is $A_V\approx 0.7\,$mag, consistent with another determination using the RC method ($A_V\approx 0.66\,$mag; Tatton et al. in prep.).

A grid element is considered a reliable control field candidate if its UMS and PMS counts, and mean extinction are within $1\sigma$ from the mean value of the respective Gaussian distributions. Usually over 100 suitable control field candidates are found within any given $1000\arcsec$ search area (61 is the minimum). The algorithm automatically selects the $N$ nearest control field candidates and combines them into a master control field. The chosen value for $N$ depends on the grid radius: $N=$10, 14, 20, 28, and 40 for the $90\arcsec$, $75\arcsec$, $60\arcsec$, $50\arcsec$, and $40\arcsec$ grids, respectively. These values provide a good balance between computing time and well sampled control field populations.
    
 \subsection{CMD residuals and significance maps}
    \label{subsec:maps}
    In order to compare the stellar content of a grid element to that of the respective master control field, it is necessary to correct for differences in the reddening, since reddened main-sequence stars can occupy the same region in the CMD as PMS stars. Using the elements' mean extinction values from Sect.\,\ref{subsec:cfields} is not appropriate since it would not account for differential reddening.
Instead the magnitudes of the stars in the control fields are individually corrected. We start by determining the cumulative colour distribution of stars in the RC box (as defined in Sec.\,\ref{subsec:cfields}) for each grid element and the corresponding master control field (Fig.\,\ref{fig:red_correction}).
For every control field star a random number is generated. This number is taken as a percentile in the cumulative distributions of the RC stars; each star is in turn {\it dereddened} by the corresponding $E(Y-K_s)$ in the colour distribution of the control field, and subsequently {\it reddened} by the corresponding $E(Y-K_s)$ in the colour distribution of the grid element. By shifting all stars from the control field along the reddening vector so that both RC colour distributions closely match each other (Fig.\,\ref{fig:red_correction}), any reddening differences between the grid element and its control field are minimized. More details on this procedure can be found in Appendix \,\ref{sec:reddeningcorrection}.

Since the reddening procedure shifts control field stars to fainter magnitudes and redder colours, its completeness values need to be adjusted. Even if the areas considered display similar extinction levels there can be differences in the completeness due to crowding.
We compare the original catalogue completeness of each control field star to the average completeness of stars located in the vicinity of its shifted CMD position (in the grid element being studied).
Control field stars are assigned a weight equal to the ratio of the original and shifted completenesses. Weights smaller than unity lead to lower densities in the Hess diagram, simulating the fact that fewer stars would have been detected.

\begin{figure}
		\resizebox{\hsize}{!}{\includegraphics{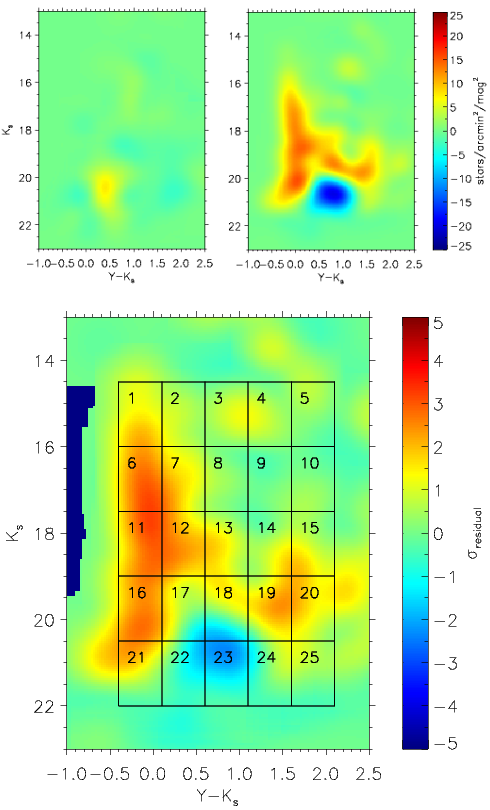}}
		\caption{\textbf{Top left:} Residual map generated for a grid element 
        showing no significant density excesses, leading to a featureless residual map; such a grid element is thus dominated by the old LMC field population. \textbf{Top right:} Residual map for a grid element
        that includes the OB association LH\,63. Significant density excesses can be seen across the CMD, due to the presence of massive OB and PMS stars. \textbf{Bottom:} Significance map for the same LH\,63 region with 25 CMD boxes overlaid; these are used to classify candidate regions based on observed density excesses (see Sec.\,\ref{sec:strategy}).}
		\label{fig:sig_maps_ex}
	\end{figure}

A Hess CMD is generated for the reddened master control field in a similar way as for the corresponding grid element (Sec.\,\ref{subsec:hess}), but using a convolution of completeness weights and Gaussian kernel in the smoothing process. Finally the Hess CMD of the reddened control field is subtracted from that of the grid element analysed. The result is a differential Hess CMD (henceforth residual map) in which differences between specific stellar populations and the local field population stand out as density excesses (examples in the top panels of Fig.\,\ref{fig:sig_maps_ex}).

We use Poisson statistics to obtain the significance of any density excesses in the residual maps. If $n_{\mathrm{ge}}$ and $n_{\mathrm{cf}}$ are the density values at a specific location in the Hess diagram for the grid element and the control field respectively, then the density excess is simply $n_{\mathrm{ge}}-n_{\mathrm{cf}}$; the individual statistical uncertainties are $\sqrt{n_{\mathrm{ge}}}$ and $\sqrt{n_{\mathrm{cf}}}$, which gives a total uncertainty of $\sqrt{n_{\mathrm{ge}}+n_{\mathrm{cf}}}$ for the residual. The statistical significance in the residual is thus:
\begin{equation}
	\label{eq:significance}
	\sigma_{\mathrm{residual}} = \frac{n_{\mathrm{ge}}-n_{\mathrm{cf}}}{\sqrt{n_{\mathrm{ge}}+n_{\mathrm{cf}}}}.
\end{equation}
This significance is computed for every point in the residual map, creating an associated significance map. Figure\,\ref{fig:sig_maps_ex} (bottom panel) shows an example significance map. We apply the algorithm to the five grids with different radii; for every element in each grid a residual map and a significance map are obtained.

\section{Identifying candidate young regions}
\label{sec:strategy}
  	
We developed a procedure that analyses the residual and significance maps, flagging and categorising candidate regions. This was extensively tested by comparing the significance maps from grid elements containing the well-studied associations LH\,60 and LH\,63, and a nearby control field \citep{Gouliermis2011}. Figure\,\ref{fig:sig_maps_ex} (top) shows residuals for the control field (left) and a grid element that contains LH\,63 (right).  The bottom panel shows the significance map which results from applying Eq.\,\ref{eq:significance} to the LH\,63 residual. It displays extended areas with significances $>2$, at CMD locations that are indicative of the presence of young UMS and PMS stars.

Also noticeable is a prominent blue patch, suggesting a field over-subtraction, where one would expect faint main-sequence stars. This feature is likely caused by small scale variations in the completeness due to crowding which are not accurately accounted for in our method. The dark blue areas at the edge of the map are artefacts caused by the absence of stars at these CMD positions for the grid element and the corresponding control field.

We divided the colour$-$magnitude space in 25 boxes (Fig.\,\ref{fig:sig_maps_ex} bottom panel), which are separately analysed.  A box is flagged when the average statistical significance is higher than a predefined threshold. The threshold value is chosen to balance sensitivity to less populous associations and robustness against statistical fluctuations. To find an appropriate threshold, we analyse the distribution of the average significances in all CMD boxes across the whole tile. This distribution is approximated by a Gaussian, and the width $\sigma$ describes the typical statistical fluctuation. We use either $2.5\sigma$ or $3\sigma$ as the flagging threshold (details to follow).

Depending on their properties (i.e. age and total  mass) stellar populations create density excesses above the local field population in different areas of the significance maps.
We analysed the residuals and significance maps for example regions with known young populations, LH\,60 and LH\,63, \citep[age 3\,$-$\,5\,Myr;][]{Gouliermis2011}.
These authors constructed catalogues of candidate PMS and UMS stars based on a statistical analysis of HST photometry in the $F\mathit{555}W$ and $F\mathit{814}W$ filters.
After correcting for a systematic difference of $0.42\arcsec$ in RA, we cross-correlate the HST and VMC catalogues
with a conservative $0.3\arcsec$ matching radius. We further compare the magnitudes of the matched pairs in VISTA's $Y$ filter and in HST's $F\mathit{814}W$ filter. The transmission curves of these filters are similar enough that these magnitudes should be comparable. For the matched pairs $(F\mathit{814}W-Y)$ is on average 0.43\,mag, with a dispersion of 0.86\,mag. To get a clean sample, pairs with $(F\mathit{814}W-Y) \geq 2$\,mag are excluded.

       \begin{figure}
		\resizebox{\hsize}{!}{\includegraphics{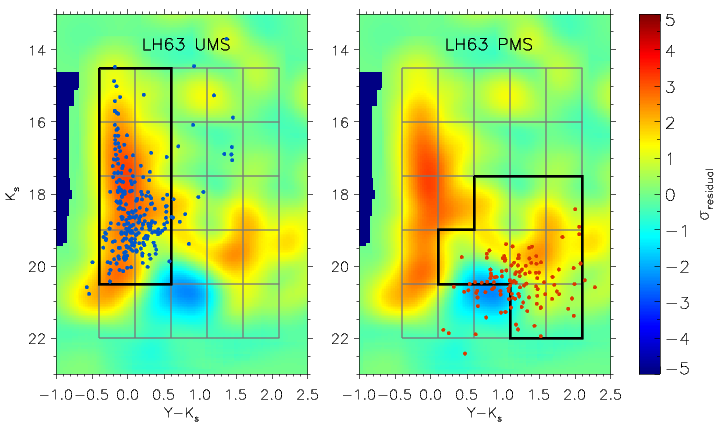}}
		\caption{The background shows the same significance map as seen in Fig. \ref{fig:sig_maps_ex} (bottom). Blue and red symbols display VMC sources that are successfully matched to the LH\,63 UMS (left) and PMS (right) catalogues from \cite{Gouliermis2011}. Thick black lines highlight the boxes relevant for the UMS and PMS classifications, respectively (see also Fig. \ref{fig:classify}).}
		\label{fig:Gou_matches}
	\end{figure}
    
    \begin{figure*}
   \centering
   \includegraphics[width=18cm]{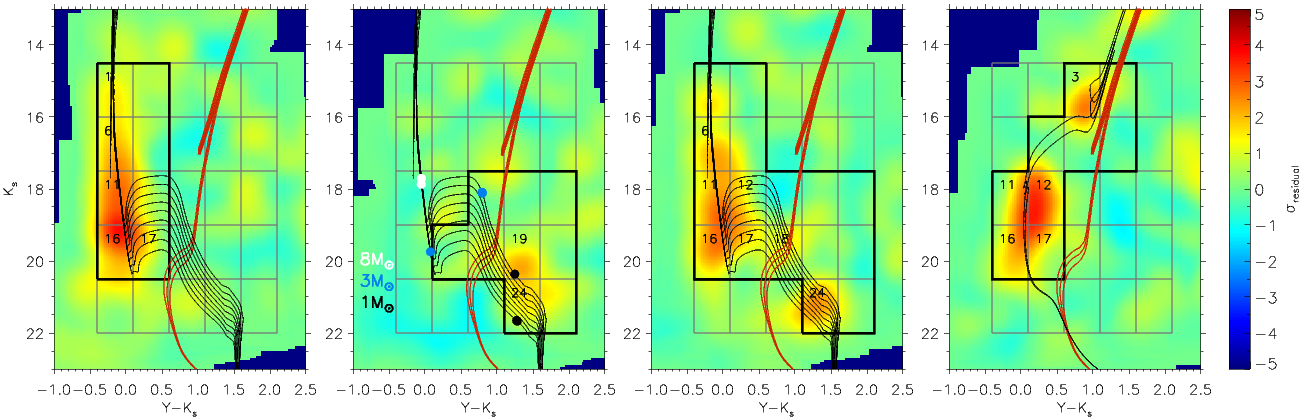}
   \caption{ \textbf{From left to right:} Significance maps for grid elements classified as UMS-only, PMS-only, UMS+PMS, and ``old''. The CMD boxes relevant for each classification are highlighted. Boxes that are flagged in the particular significance map are numbered. PARSEC isochrones \citep{Bressan2012} are shown in all panels. Red isochrones represent ages from $\log{(t/\mathrm{yr})} = 9.6$ to 9.8, with a metallicity of $Z=0.0033$ \citep{Tatton2013}; they show the typical location of the old LMC field population. Black isochrones represent young populations from $\log{(t/\mathrm{yr})} = 6.0$ to 6.8 ($Z=0.008$; \citealt{Rubele2012}). The two black isochrones in the rightmost panel represent populations of $\log{(t/\mathrm{yr})} = 8.5$ and 8.6 ($Z=0.008$). All isochrones are reddened according to the mean extinction for that particular grid element (Sect.\,\ref{subsec:cfields}). In the second panel the theoretical positions for stars of three different masses are shown (colour-coded circles) for the youngest and oldest black isochrones.}
   \label{fig:classify}%
\end{figure*}

In LH\,60 we find 112 and 174 VMC counterparts to HST's PMS and the UMS sources, respectively. For LH\,63 the corresponding numbers are 125 and 269 \footnotemark \footnotetext{Based on a catalogue matching with shifted coordinates, we estimate that $\sim 3$\% of the UMS matches and $\sim10$\% of the PMS matches are spurious.}.  Figure \ref{fig:Gou_matches} shows the VMC counterparts for the LH\,63 UMS and PMS catalogues, with the significance map shown in the background. A clear gap is visible between both populations, since the HST catalogue excludes areas in the optical CMD which are heavily contaminated by field stars. Clearly, the UMS matches coincide very well with areas of statistically significant density excesses; the highlighted CMD boxes (Fig.\,\ref{fig:Gou_matches} left) cover the majority of UMS matches and the corresponding density excesses. Overall, 237 out of 269 UMS matches are located within these eight boxes; some matches are located near the red giant branch, and are likely contaminants in the UMS catalogue \citep{Gouliermis2011}.

The situation is more complicated for the PMS matches. Some matches fall on the strong negative density excess described previously; we did not use this part of the CMD to identify PMS populations precisely to avoid severe contamination by main-sequence stars.
The large scatter in the CMD distribution of the PMS matches can be due to photometric errors and/or young stellar variability (e.g. T\,Tauri stars or FU\,Orionis variables; \citealt{Contreras2014, Rice2015}). In addition, the PMS population could have either an age spread, or it could consist of multiple populations with different ages, similar to what has been seen in Orion \citep{Beccari2017}.  Nevertheless, the regions with the most significant density excesses suggest a PMS distribution that is brighter than the distribution of HST-matched stars. Two effects likely contribute to this. Firstly, small-scale completeness variations can be significant in crowded regions like LH\,63 (see Sec. \ref{subsec:maps}). This can lead to over-subtraction during the control field decontamination process which eliminates a potential density excess due to faint PMS stars.
Secondly, the HST PMS catalogue included mostly relatively faint PMS stars ($F\mathit{814}W \geq 21.2$\,mag). The density excess seen in our maps includes brighter PMS candidates excluded from HST optical catalogues.
Using the location of the PMS matches and the location of the typical density excesses seen in the residual maps of known young associations, we selected the CMD boxes highlighted in the right panel of Fig.\,\ref{fig:Gou_matches} as PMS indicators. 96 out of 125 PMS matches are within this area and overall around 85\,\% of the HST$-$VMC counterparts are located within the outlined UMS and PMS boxes.

Having identified contiguous regions in the CMD indicative of the presence of specific populations, we classify each grid element based on which boxes are flagged in its significance map.
Four classifications are adopted:

\begin{enumerate}
\item \textbf{UMS-only:} At least two adjacent boxes are flagged in the UMS CMD region. Figure \ref{fig:classify} (left) shows a typical example of this classification. It covers a broad age range from $\sim$\,10 to $\sim 300\,$Myr. Based on artificial cluster tests (see Sec. \ref{subsec:synthetic_clusters}) a minimum cluster mass of $\sim 500\,\mathrm{M_{\sun}}$ is needed to reliably flag two UMS boxes. For populations younger than $10\,$Myr the PMS population is also detectable, changing its classification to UMS+PMS (see below). Beyond $\sim300\,$Myr sufficiently massive clusters ($> 1000\,\mathrm{M_{\sun}}$) create a significant red giant excess, which will trigger the classification ``old''.

\item \textbf{PMS-only:} At least two adjacent boxes are flagged in the PMS CMD region (second panel in Fig.\,\ref{fig:classify}). This classification traces young ($<10\,$\,Myr), low-mass clusters and associations up to around $\sim 500\,\mathrm{M_{\sun}}$. While the PMS phase for stars with masses $\lesssim 0.5\,\mathrm{M_{\sun}}$ can last up to 100\,Myr \citep[e.g.][]{Tout1999}, PMS populations older than $\sim 10$\,Myr are too faint to be detected in the VMC survey. Young clusters and associations with masses above 500\,$\mathrm{M_{\sun}}$ flag also boxes typical of UMS populations, changing the classification to UMS+PMS.

\item \textbf{UMS+PMS:} A grid element is classified as UMS+PMS if a total of at least three boxes in both the typical PMS and UMS CMD regions are flagged. Adjacency is not strictly enforced, since a minimum of three flagged boxes always leads to reasonable combinations. The same age range as in the PMS classification is probed ($<10\,$Myr), but the clusters and associations are more massive, since enough massive stars need to be present to flag UMS boxes. Examples for this classification are shown in Fig.\,\ref{fig:classify} (third panel) and Fig.\,\ref{fig:sig_maps_ex}.

\item \textbf{Old:} This classification requires a minimum of three flagged boxes in the red giant branch and the fainter parts of the UMS CMD region. The flagging threshold for the red giant branch boxes is $2.5\sigma$ (compared to $3\sigma$ for the other boxes). We found in tests with synthetic and real clusters that this threshold reduction improves our ability to classify old clusters, without a noticeable increase in the number of false positives. Grid elements with clusters older than $\sim300\,$Myr and more massive than $\sim 1000\,\mathrm{M_{\sun}}$ fall into this classification (Fig.\,\ref{fig:classify}, right).
\end{enumerate}

\section{Testing the identification strategy}
\label{sec:testing}
\subsection{Synthetic clusters}
\label{subsec:synthetic_clusters}
To assess the procedure's sensitivity to young populations of different masses and ages we ran tests using synthetic clusters.
To generate the synthetic clusters we used the Popstar Evolutionary Synthesis Code \citep{Molla2009} and adopted a Kroupa IMF \citep{Kroupa2001,Kroupa2002}. There is no conclusive evidence that the IMF in the LMC is significantly different from the Galactic IMF \citep[for $M > 1\,\mathrm{M_{\sun}}$; ][]{Gruendl2009,Liu2009a,Liu2009b}, with the possible exception of 30\,Doradus \citep{Schneider2018}. NIR photometry for the VISTA filter set was obtained using PARSEC models \citep{Bressan2012}. Synthetic clusters were generated for the following mass and age combinations:
\begin{itemize}
\item $M_{\mathrm{cl}} \in [250\,\mathrm{M_{\sun}},\,500\,\mathrm{M_{\sun}},\, 1000\,\mathrm{M_{\sun}},\,2000\,\mathrm{M_{\sun}},\,3000\,\mathrm{M_{\sun}}]$
\item $\log{(t/\mathrm{yr})} \in [6.0,\,6.3,\,6.7,\,7.0,\,7.5,\,8.0,\,8.5,\,9.0]$.
\end{itemize}
The cluster masses are representative of LMC clusters within this age range \citep{Grijs2006}. We also adopted the canonical LMC metallicity of $Z=0.008$. For every mass$-$age combination ten clusters were created. Incompleteness was applied and photometric errors added to match the quality of the VMC data before injecting the clusters into the PSF tile catalogue. All clusters were seeded in control field like grid-elements with flat significance maps. Each synthetic cluster is fully contained within a grid element, a reasonable assumption given that even the smallest elements have a physical radius of 10\,pc (1\,pc $\cong 4\arcsec$ at the LMC distance). The enhanced PSF tile catalogues for every synthetic cluster are ingested into our algorithm, and the resulting residuals and significance maps are evaluated. If a synthetic cluster is classified into one of the four classes defined in Sec. \ref{sec:strategy} and in agreement with the cluster input properties, it is considered to be reliably identified.

\begin{figure}
		\resizebox{\hsize}{!}{\includegraphics{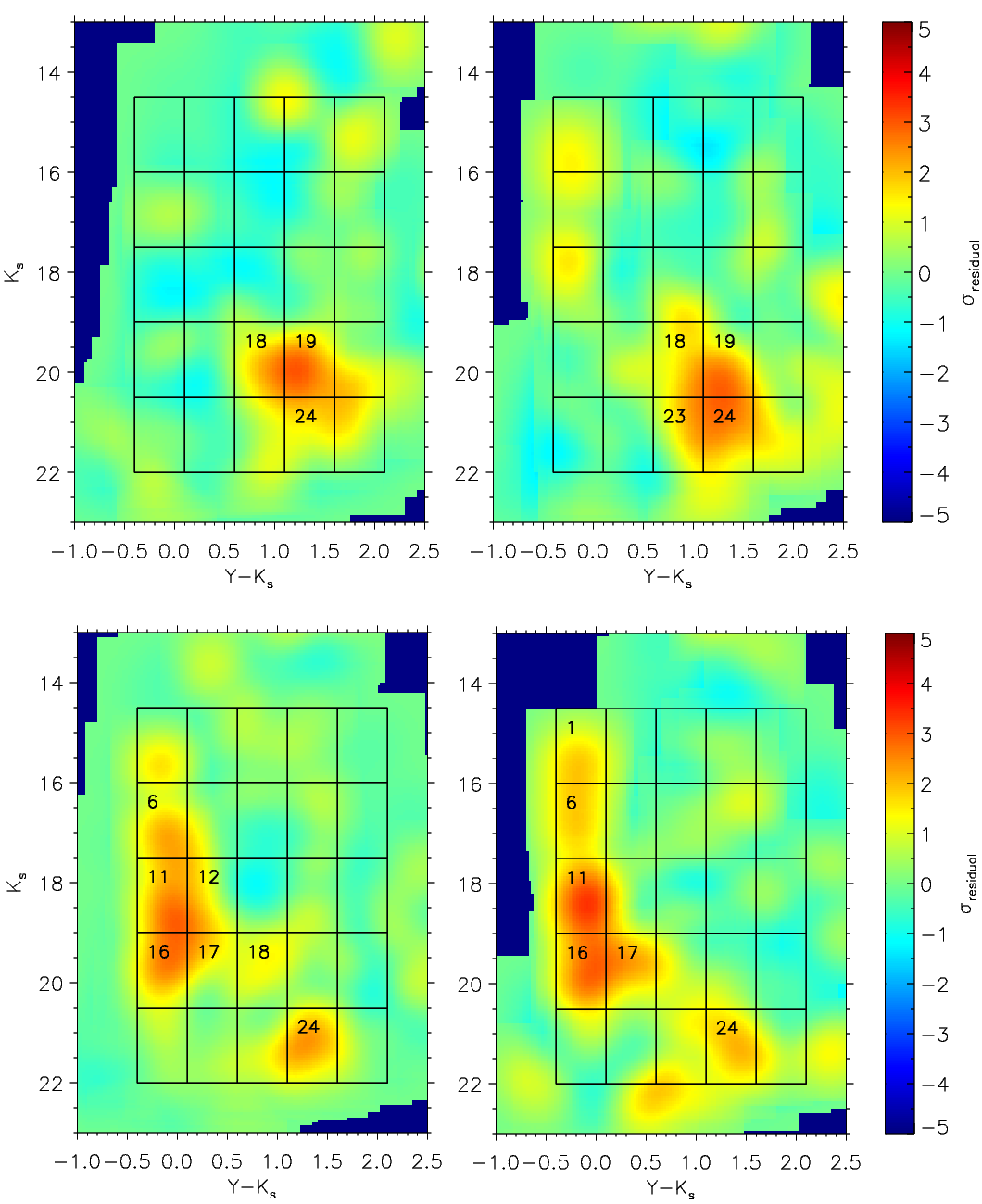}}
		\caption{\textbf{Top:} The significance map for a PMS-only classified grid element (left) is compared 
to a synthetic stellar population of $250\,\mathrm{M_{\sun}}$ and 1\,Myr (right). \textbf{Bottom:} Similar to the top, but showing a UMS+PMS classified grid element (left)
and a synthetic stellar population of $1000\,\mathrm{M_{\sun}}$ and 5\,Myr (right). Flagged boxes are numbered in each map.}  
		\label{fig:real_art_com}
	\end{figure}

In Fig. \ref{fig:real_art_com} we show examples of significance maps generated from observed data, compared with maps generated from synthetic clusters. The top left panel shows a PMS-only classified element with a clear PMS signal and no UMS excess. This suggests the presence of a very young, low mass population.  Indeed a synthetic population of $250\,\mathrm{M_{\sun}}$ with an age of 1\,Myr generates a very similar significance map. The bottom left panel contains the significance map for a UMS+PMS classified element. A synthetic stellar population of $1000\,\mathrm{M_{\sun}}$ with an age of 5\,Myr creates a similar significance map. Since at 5\,Myr many PMS stars fall below the sensitivity limit, the PMS signature is relatively weak.

\begin{figure}
		\resizebox{\hsize}{!}{\includegraphics{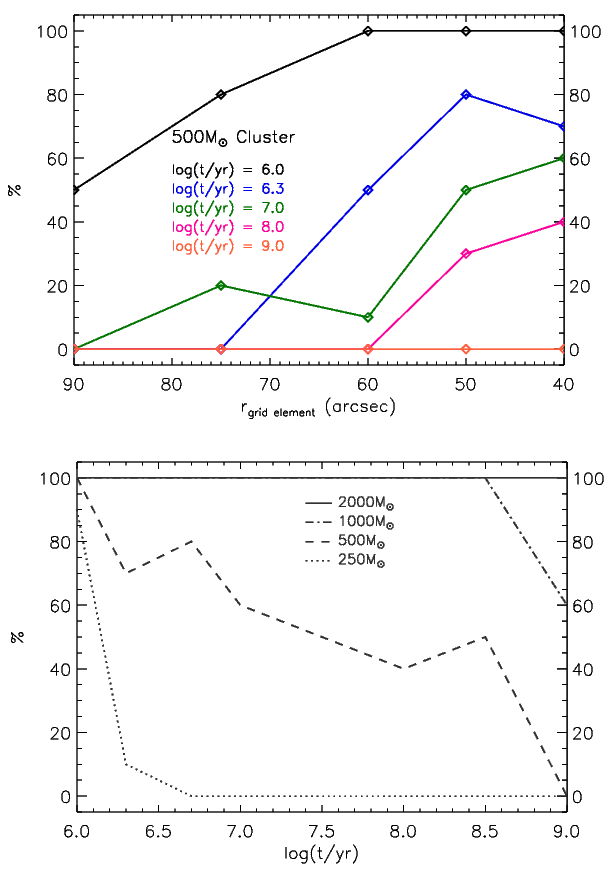}}
		\caption{\textbf{Top panel:} Detection percentages for $500\,\mathrm{M_{\sun}}$ clusters of different ages for grid-element radii from $90\arcsec$ to $40\arcsec$. A general increase in detection rates with decreasing radius is noticeable. Also apparent is a decrease in detection rates with increasing cluster age. \textbf{Bottom panel:} Detection percentages for clusters of four different masses across the age range from 1\,Myr to 1\,Gyr for the $40\arcsec$ radius. A decrease in sensitivity for older ages and lower masses is apparent. Clusters more massive than $500\,\mathrm{M_{\sun}}$ show high detection rates for young ages.}
		\label{fig:detection_rates}
	\end{figure}

Fig.\,\ref{fig:detection_rates} shows the results of our synthetic cluster tests. In the top panel the detection rates for the different grid element radii for $500\,\mathrm{M_{\sun}}$ clusters of various ages are presented. While for the $90\arcsec$ radius clusters of this mass are rarely detected, the percentages increase steadily for smaller radii. At a radius of $40\arcsec$ the detection rates are, for ages where one would expect to find PMS stars, at least 60\,\%. The bottom panel looks more closely at the results obtained using the $40\arcsec$ grid for four different cluster masses. For masses $\geq 1000\,\mathrm{M_{\sun}}$, clusters are always detected with a drop in detection rate only noticeable at 1\,Gyr.
For lower masses the detection rate drops steadily with age.
This is due to a decrease of flagged UMS boxes with increasing age, as more massive stars evolve away from the main sequence onto the red giant branch. At these cluster masses this does not necessarily trigger the flagging of boxes that lead to the ``old'' classification. For ages $\lesssim 10\,$Myr the majority of $500\,\mathrm{M_{\sun}}$ clusters are detected. Even though the detection rate drops sharply for masses $<500\,\mathrm{M_{\sun}}$, it remains high for very young ages.

These tests reveal three clear trends. Rather obviously, the more massive a synthetic cluster is at any given age, the higher the probability of a reliable detection.  Secondly, with smaller grid element radius the probability of detection for a given synthetic cluster mass and age increases.
The reason is that the same number of synthetic stars cause a higher density excess in the residuals of smaller grid elements, leading to a higher flagging probability.
Thirdly, the detection rates generally decrease with age. The luminosity of intermediate/low mass PMS stars decreases as they approach the main sequence.  Hence, an increasing fraction of PMS stars falls below the sensitivity limit of the VMC survey for progressively older ages.

\begin{table*}
\caption{Most likely classifications for mass and age ranges based on synthetic cluster tests.}    
\label{table:classifications}     
\centering 
\begin{tabular}{c c c c c c}        
\hline\hline 
 & $<2\,$Myr & 2 -- 5\,Myr & 5 -- 10\,Myr & 10 -- 300\,Myr & $>300$\,Myr  \\    
\hline
$<500\,\mathrm{M_{\sun}}$        & PMS-only    & - & - & - & - \\ 
$500 - 1000\,\mathrm{M_{\sun}}$  & UMS+PMS & UMS+PMS & UMS-only & UMS-only & UMS-only \\
$>1000\,\mathrm{M_{\sun}}$       & UMS+PMS & UMS+PMS & UMS+PMS & UMS-only & old \\ 
\hline  
\end{tabular}
\end{table*}

Table\,\ref{table:classifications} presents an overview of the most common classifications for synthetic clusters of different age and mass ranges. The decrease of detectable PMS stars with age causes an increase in the minimum cluster mass necessary to detect a PMS signature. Beyond 10\,Myr the VMC survey is not sufficiently deep to reliably detect any remaining PMS stars. Therefore, all detected clusters are classified either as UMS-only or as ``old''.
The tests reveal that massive ($\gtrsim 2000\,\mathrm{M_{\sun}}$) clusters within the age range 30\,Myr -- 1\,Gyr can be mis-classified as UMS+PMS, and thus contaminate the UMS+PMS class. However, the level of contamination is 2.5\% at most. 

In summary, we conclude that, for a given mass, a cluster or association is easier to detect and classify at a young age (preferably $< 10$\,Myr); this is mostly due to the sensitivity and completeness limits of the VMC survey.  While $\sim 28$\,\% of detected clusters with age 10\,Myr have a PMS signature leading to a PMS-only or UMS+PMS classification, this fraction increases to $\sim 66$\,\% for 5\,Myr-old populations and $\sim 98$\,\% for very young ages  ($\leqslant 2$\,Myr). More massive clusters are obviously more likely to be identified. Another important result is that finer grids are better suited to find low-mass clusters, despite the effects of small number statistics due to the decreasing number of stars per grid element.

\subsection{Literature clusters}
\label{subsec:literature_clusters}
The synthetic cluster tests provided valuable results about the mass and age ranges that are effectively traced with our method. However, as these synthetic clusters were only injected into the PSF catalogue rather than into the images, information on the sensitivity to different cluster radii and star count density profiles is lacking. We compiled a list of 31 clusters and associations from the literature \citep{Gouliermis2003,Glatt2010,Popescu2012}. Our list is not a complete census of clusters in the pilot field, but provides a reliable sample of clusters with different ages and sizes. The selected systems span an age range from a few Myr up to around 1 Gyr, and apparent sizes from $10\arcsec$ to over $100\arcsec$ (sizes were estimated  visually from the VMC images).

Table \ref{table:detections} shows how many clusters are classified as a function of grid element radius.  With decreasing grid element size the number of unclassified clusters decreases monotonically. This trend is in line with the results from the synthetic cluster tests where an increasing sensitivity for the finer grids was observed. For the finest grid 25 out of 31 literature clusters are classified with our method. 
For 23 of these, their classifications and inferred broad age range (see Tab.\,\ref{table:classifications}) are consistent with published literature ages. The two remaining clusters have literature ages of $\log{(t/\mathrm{yr})}\sim$\,7.8; since a RC signature is detected, our method classifies these clusters as old, implying an age $\gtrsim 300$\,Myr.

Six unclassified clusters remain, four of which flag a single CMD box. Since this does not trigger a classification, these four clusters are at the sensitivity limit of the $40\arcsec$ grid.
The unclassified clusters are either spatially small compared with the grid element and/or relatively old (based on their literature ages). Two of the six unclassified clusters are the smallest in our list with radii of $10\arcsec$ and $14\arcsec$, $\sim6$\% and $\sim$12\% of a grid element area. This indicates that our method is mostly sensitive to clusters with $r\gtrsim3\,\rm{pc}$ at the LMC distance. For four of the six unclassified clusters the literature age is in the range $7.3 \lesssim \log{(t/\mathrm{yr})} \lesssim 8.7$. The stellar populations of old clusters move towards areas in the CMD which are more heavily contaminated by the old field population, further decreasing the sensitivity. This confirms the results from the synthetic cluster tests that comparatively old systems have lower detection rates. Given that our stated goal is to identify young populations, a decrease of detection rate with cluster age is not an issue.

\subsection{Final choice of grid element radius}
\label{subsec:final_choice}
Our analysis clearly advocates the use of the $40\arcsec$ radius grid because of its increased sensitivity. A further decrease in radius leads to grid elements without any RC stars,  impairing the methods's ability to correct for reddening differences between a grid element and its control field. On average, a $40\arcsec$ grid element is populated by 265 stars, with 171 stars being the minimum. Our subsequent analysis focuses on the optimal $40\arcsec$ radius grid.

\begin{table}
\caption{Number of classified clusters from the sample of 31 clusters from the literature, as a function of grid element radius.} 
\label{table:detections}  
\centering     
\begin{tabular}{c c c c c c}
\hline\hline 
 & $90\arcsec$ & $75\arcsec$ & $60\arcsec$ & $50\arcsec$ & $40\arcsec$ \\ 
\hline
classified  & 20 & 23 & 23 & 24 & 25 \\ 
unclassified  & 11 & 8 & 8 & 7 & 6 \\
\hline
\end{tabular}
\end{table}

\section{Results}
\label{sec:results}
Applying our method to the PSF catalogue from the pilot field provides the residual and significance maps, flagged boxes, and classification for every grid element. For the $40\arcsec$ grid, 10,730 grid elements are unclassified, 298 are classified as UMS-only, 84 as PMS-only, 124 as UMS+PMS, and finally 14 are classified as ``old''.

\subsection{Spatial distribution of the young populations}
\label{subsec:spa_distribution}
\subsubsection{Global properties}
\label{subsubsec:global_spatial}
\begin{figure*}
   \centering
   \includegraphics[width=17cm]{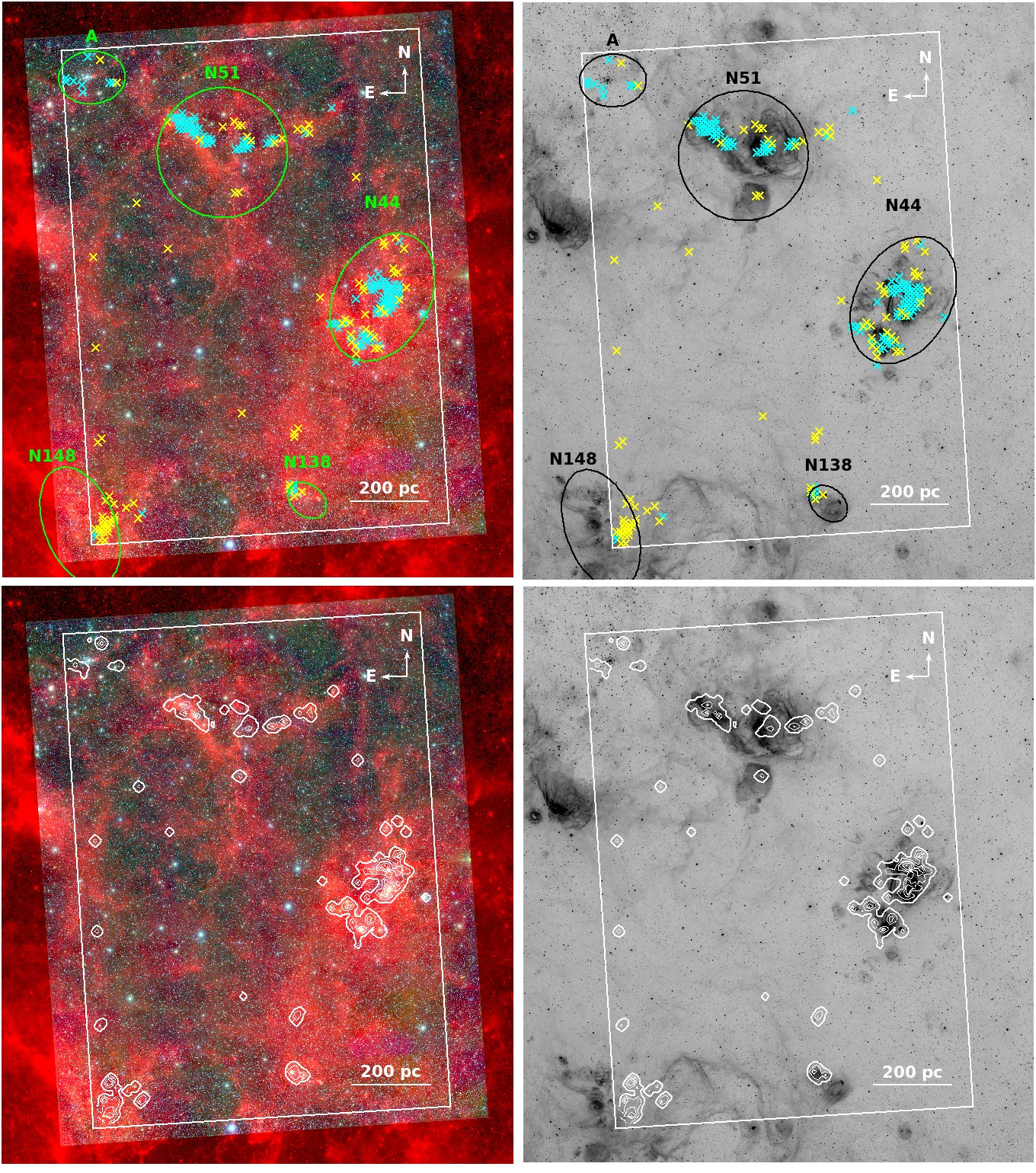}
   \caption{\textbf{Top left:} Three-colour composite image with VMC $Y$ (blue) and $K_s$ bands (green), and \textit{Spitzer} IRAC $8.0\,\mathrm{\mu m}$ (red). The rectangle shows the region covered by our analysis. Small crosses mark the centres of grid elements for the $40\arcsec$ grid which are classified as PMS-only (yellow) or UMS+PMS (cyan).  Several prominent regions are highlighted and labelled \citep[spatial sizes according to][]{Bica2008} and discussed further in Sec.\,\ref{sec:results}. \textbf{Top right:} Inverted grey scale map of $\mathrm{H\alpha}$ emission \citep{Smith2005}. \textbf{Bottom:} PMS density contours for elements classified as PMS-only or UMS+PMS derived from the residual maps of the grid elements. The outermost contour represents $\Delta n_{\mathrm{PMS}} = 2.4\,\textrm{stars arcmin}^{-2}$ (see Sec.\,\ref{subsec:PMS_count} for details); every subsequent contour represents an increase in density by $3\times \Delta n_{\mathrm{PMS}}$.}
   \label{fig:PMS_dist_class}%
\end{figure*}

    \begin{figure}
		\resizebox{\hsize}{!}{\includegraphics{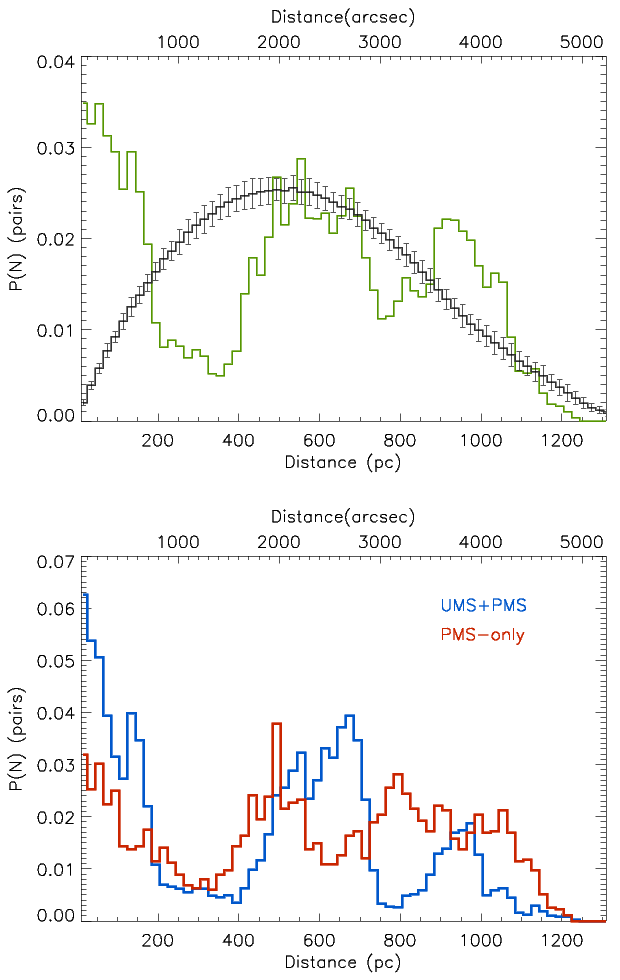}}
		\caption{\textbf{Top:} Normalized distance distribution for all possible grid pairs for the 208 PMS-only and UMS+PMS classified elements (green line), and for a same size sample of randomly distributed unclassified elements (black line). \textbf{Bottom:} As the top panel, but separating the two classifications.} 
		\label{fig:TPCF_grid}
	\end{figure}

Figure\,\ref{fig:PMS_dist_class} shows the spatial distribution of the 208 grid elements with a significant PMS contribution (classified as UMS+PMS and PMS-only, cyan and yellow crosses, respectively).
These grid elements are not distributed uniformly but instead concentrate in areas with enhanced dust emission as traced by \textit{Spitzer} in the IRAC $8.0\,\mathrm{\mu m}$ band \citep{Meixner2006}. Around 80\% of them are found in three major regions, namely N\,44, N\,51, and N\,148 (see Fig.\,\ref{fig:zoom_complexes} for a detailed view). N\,44 and N\,51 are well-studied large star formation complexes \citep[e.g.][]{Carlson2012}. Almost all classified elements associated with these complexes are located within regions of about 290 $\mathrm{arcmin}^2$ ($\sim 65,200\,\mathrm{pc^2}$) and 380\,$\mathrm{arcmin}^2$ ($\sim 85,500\,\mathrm{pc^2}$), respectively. On these spatial scales star formation complexes contain young populations formed in multiple and/or extended star formation events \citep[see][for a detailed review]{Gouliermis2018} For comparison, the HTTP survey of the 30\,Doradus region covers $\sim 168\,\mathrm{arcmin^2}$ \citep{Sabbi2013}, within which \cite{Schneider2018b} found evidence for complex spatial and temporal substructure amongst the massive stars. To the south-eastern corner of the field there is another concentration of classified elements associated with N\,148, which is also prominent in CO \citep{Wong2011} and dust emission \citep{Meixner2006}. A fourth group is situated to the North$-$East (labelled Region\,A). While the numbers of PMS-only and UMS+PMS classified grid elements are small, this region exhibits the largest concentration of UMS-only classified elements (see Appendix \ref{sec:dist_old}), indicating a comparatively mature population. Another concentration of classified elements is associated with the emission nebula N\,138. A more detailed discussion of the individual regions mentioned above is found in Sec.\,\ref{subsubsec:regions}. 

Outside these complexes, classified grid elements are fairly scattered; as can be seen in Fig.\,\ref{fig:PMS_dist_class} there are 12 isolated elements with signatures of a young population. An inspection of the residuals and significance maps for these isolated elements shows that they are generally consistent with the maps from elements located within star forming complexes and from the synthetic clusters. A comprehensive analysis of the stellar populations of these isolated elements is beyond the scope of this paper.

Overall, UMS+PMS classified grid elements are almost exclusively found in groups. PMS-only grid elements, on the other hand, are often located on the outskirts of populous UMS+PMS star groups or appear isolated. To quantify the degree of clustering we calculated the distance between every possible pair of classified elements. Figure \ref{fig:TPCF_grid} (top) shows this distance distribution, together with a distribution for a same size sample of randomly placed unclassified elements.
To eliminate statistical fluctuations we ran this simulation 100 times and use the mean and standard deviation values to construct the random distribution histogram with corresponding uncertainties. Due to the finite pilot field size the number of possible random pairs decreases for large distances. For small distances ($<180\,$pc) the observed distribution shows a clear excess of classified pairs as a consequence of the clustered distribution of young stars. The broad peak between 450\,pc and 700\,pc is due to the distances between N51 and N44, and between N51 and Region A. The peak between 900\,pc and 1050\,pc is caused by the distances between N44 and N\,148, N44 and Region\,A, and N51 and N\,148.

Figure \ref{fig:TPCF_grid} (bottom) displays the distance distribution for the PMS-only and UMS+PMS subsamples separately. They are distinct and not simply scaled-down from the overall distribution. As a result of the very strong clustering of the UMS+PMS classified elements, the corresponding distribution shows very prominent peaks. The very narrow peak between $\sim 130$\,pc and $\sim 170$\,pc is due to the distances between associations within N44 and N51. The broad peak for distances $\lesssim 200$\,pc suggests a similar distribution of UMS structure sizes as seen in other star formation complexes in the LMC \citep{Sun2017a,Sun2017b} and in the SMC \citep{Sun2018}.

In contrast, the PMS-only distribution shows smaller variations, in agreement with a more extended spatial distribution. It is a common observation in young clusters that high mass stars are more centrally located than intermediate/low mass stars \citep[e.g.][]{Zinnecker1993,Gennaro2011,Pang2013}. The underlying physical process is assumed to be dynamical mass segregation \citep[e.g.][]{Bonnell1998, Allison2009}.
Such effects are relevant within individual clusters, which are usually smaller than one element in our grid. However, since segregation time-scales are on the order of several crossing times \citep[e.g.][]{Grijs2002}, dynamical mass segregation is too slow to explain the observed spatial distribution on the scales of the complexes N\,44 and N\,51.

In agreement with our results, the HTTP survey found that also for the 30\,Doradus region the UMS stars mostly concentrate in a few major population centres \citep{Sabbi2016}, while the PMS stellar distribution displays a larger spatial dispersion \citep{Cignoni2015}.  These authors see evidence of constructive feedback from massive stars igniting the birth of new generations of stars. In N\,44, \citet{Chen2009} also report evidence of triggered star formation. Triggering by massive stars could operate on larger spatial scales than mass segregation; it is thus a viable scenario for the different spatial distribution between PMS-only and UMS+PMS elements, since it could lead to the formation of less massive clusters or associations in the outskirts. This is observed in the Carina Nebula, where the currently ongoing star formation seems to produce only stars up to $\sim 10\,\mathrm{M_{\sun}}$ \citep{Gaczkowski2013}.

Alternatively, the evaporation of bound clusters due to gas expulsion could also result in the observed spatial distribution of classified elements. After a quick gas removal phase ($\lesssim 1\,$Myr), clusters are predicted to expand to half mass radii $\gtrsim 10$\,pc (for masses $\sim 5000\,\mathrm{M_{\sun}}$) within 10\,Myr \citep{Pfalzner2014}. Crucially they  develop an extended stellar halo of ejected stars which can extend beyond 100\,pc \citep{Moeckel2010}. In the comparatively low stellar density environment of such halos there will naturally be fewer UMS stars, thus increasing the likelihood of PMS-only classified elements in the cluster outskirts.

\subsubsection{Individual regions}
\label{subsubsec:regions}
Several regions, associated with a large number of classified elements, are identified in Fig.\,\ref{fig:PMS_dist_class}. As already mentioned, in  N\,51 and N\,44, the UMS+PMS classified elements are strongly clustered whilst the PMS-only elements are often located in the outskirts of those regions. In both complexes significant dust and $\mathrm{H\alpha}$ emission is also observed. In N\,44, $\mathrm{H\alpha}$ emission and the UMS+PMS classified elements are spatially coincident, as would be expected since the ionising radiation from the UMS stars is the origin of the $\mathrm{H\alpha}$ emission. Two large substructures can be seen: the larger one near the centre corresponds to the associations LH\,47 and LH\,48, and the smaller one towards the South includes LH\,49. To the East of  N\,51, a similarly strong overlap between $\mathrm{H\alpha}$ emission and classified elements is seen. To the West the intense $\mathrm{H\alpha}$ emission traces a bubble-like shape, but the overlap with classified elements is patchy. Two OB associations, LH\,51 and LH\,54, are associated with this bubble \citep[e.g.][]{Book2009}.

Besides these complexes the two most prominent concentrations of classified elements are located in the western part of N\,148 and in Region A. N\,148 is an intense and vigorous star forming region \citep[e.g.][]{Ambrocio2016}, only partially covered by our analysis. Most grid elements are classified as PMS-only without a significant UMS population; this suggests the existence of a distributed population of intermediate/low mass PMS stars associated with N\,148. This population has however not been previously identified \citep[][and references therein]{Bica2008}.
Based on comparisons of the significance maps with synthetic clusters, some of the PMS-only grid elements seem very young ($\sim 1$\,Myr). The presence of significant dust and CO emission combined with the lack of extended $\mathrm{H\alpha}$ emission provides further evidence of a young population devoid of UMS stars.
 
The classified elements in Region A combine relatively weak PMS signatures with strong UMS signatures, suggesting an older age.  The most prominent cluster in this area, NGC\,2004, has indeed an age of $\sim 20$\,Myr \citep{Niederhofer2015}, resulting in a PMS population too faint to be reliably detected with the VMC data.
Region A also lies in an area with no significant dust or H$\alpha$ emission. This all hints at a comparatively old age of the dominant populations in the region, probably a result of the energetic feedback from massive stars that have significantly eroded the interstellar medium.

Another grouping of classified elements is found to the North East of the emission nebula N\,138. More specifically the PMS-only and UMS+PMS classified regions are spatially coincident with the H\,{\sc ii} regions N\,138A and N\,138C. In N\,138A \cite{Indebetouw2004} found an ultracompact H\,{\sc ii} source, indicative of very young massive stars.

Our method to classify young populations not only identifies all major star forming complexes in tile LMC\,7\_5, but it also exposes their full extents and distribution for the first time.

\subsection{Quantitative analysis of the PMS populations}
\label{subsec:PMS_count}
Using the residual maps we calculate the number density of PMS candidates as well as the overall number of PMS candidates in the classified elements. Taking the mean density excess of the flagged boxes relevant for PMS populations (see Fig.\,\ref{fig:classify}) and multiplying it by the area covered by these boxes in the CMD gives the PMS number density. We obtain a mean PMS number density $n_{\mathrm{PMS}} =12.7\,\textrm{stars arcmin}^{-2}$ over all elements classified as PMS-only or UMS+PMS.  The uncertainty, estimated by analysing the density fluctuations in the residuals of non-classified elements, is $\Delta n_{\mathrm{PMS}}=2.4\,\textrm{stars arcmin}^{-2}$. Multiplying the PMS density by the solid angle of the grid element ($\sim 1.4\,\mathrm{arcmin^2}$) gives the number of PMS candidates \textbf{($N_{\mathrm{PMS}}$)}.
On average there are $N_{\mathrm{PMS}}=17.7\pm 3.4$ per classified element. 

In Fig.\,\ref{fig:PMS_dist_class} the PMS density contours are plotted. The highest density ($40\,\textrm{stars arcmin}^{-2}\cong 0.18\,\textrm{stars pc}^{-2}$) is found
within N\,44. Large complexes contain multiple high density peaks, displaying a hierarchical structure similar to that found for UMS stars by \citet{Sun2017b,Sun2017a}. Integrating over all classified elements and accounting for the overlap between neighbouring grid elements leads to a total number of PMS candidates of $2256 \pm 54$. This result is a lower limit due to significant incompleteness in the VMC data at magnitudes typical of PMS stars.
\begin{table}
\caption{Number of PMS candidates in the whole LMC 7\_5 tile and individual prominent regions. The errors are calculated assuming grid elements are statistically independent.}   
\label{table:3} 
\centering           
\begin{tabular}{l l}      
\hline\hline             
 & $\mathrm{N_{PMS}}$ \\ 
\hline     
   LMC 7\_5\,\, &  \llap{22}56 $\pm$ 54  \\  
   N\,44 &  \llap{10}00 $\pm$ 38 \\
   N\,51 &  $\llap{3}79\pm 22$ \\
   N\,148 &  $\llap{2}38\pm 20$ \\
   A &  $\llap{1}12\pm 13$ \\ 
   N\,138 &  $74\pm 9$ \\

\hline 
\end{tabular}
\end{table}
Table \ref{table:3} lists the number of PMS candidates for  the whole pilot field as well as for the five regions described in the previous sections. Overall, $\sim 80$\% of all PMS candidates identified are located in one of these regions, with N\,44 being the most populous.

For comparison, \cite{Meingast2016} estimated the entire young stellar population of the Orion A molecular cloud to have between 2300 and 3000 stars. Using a Kroupa IMF this gives 300 to 390 stars with masses $1\,\mathrm{M_{\sun}} \leqslant M_* \leqslant 4\,\mathrm{M_{\sun}}$, which is the PMS mass range our method is sensitive to (see Fig.\,\ref{fig:classify}). The area covered by the Orion survey (18.3\,$\mathrm{deg^2}$) corresponds to $4.5\,\mathrm{arcmin^2}$ or 3.2 grid elements at the LMC distance. In the Carina Nebula Complex, 8781 young stars were identified based on their NIR colour excess \citep{Zeidler2016}. Applying a Kroupa IMF to the same mass range gives $\sim$\,1150 stars that could potentially be identified as PMS with our method. This is comparable with our PMS count for the N\,44 complex. Note however that this estimate includes only sources with a NIR colour excess. At the LMC distance the area observed in \cite{Zeidler2016} corresponds to $51.5\,\mathrm{arcmin^2}$ or around 37 grid elements, which is approximately the area covered by the large group of classified elements in the centre of N\,44.  Incompleteness and crowding would obviously reduce the number of PMS sources we would be able to detect.

    \begin{figure}
		\resizebox{\hsize}{!}{\includegraphics{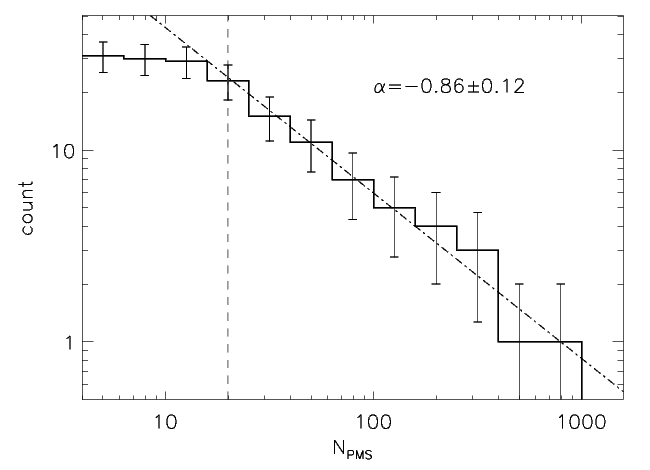}}
		\caption{Cumulative number distribution for the 31 PMS structures (0.2\,dex bins). The vertical dashed line indicates the sensitivity limit, while the dashed-dotted line represents a power-law with a slope of $-0.86$. The error bars represent the Poissonian uncertainties.}
		\label{fig:hist_counts}
	\end{figure}
To obtain a more detailed view of the morphology of the PMS populations we define PMS structures as regions enclosed by the lowest density contour in Fig. \ref{fig:PMS_dist_class}. We detect 31 structures in total, the most populous of which is located in N\,44 and contains $\sim670$ PMS candidates. Figure \ref{fig:hist_counts} shows the cumulative $N_{\mathrm{PMS}}$ distribution for the PMS structures. 
 \begin{figure*}
   \centering
   \includegraphics[width=18cm]{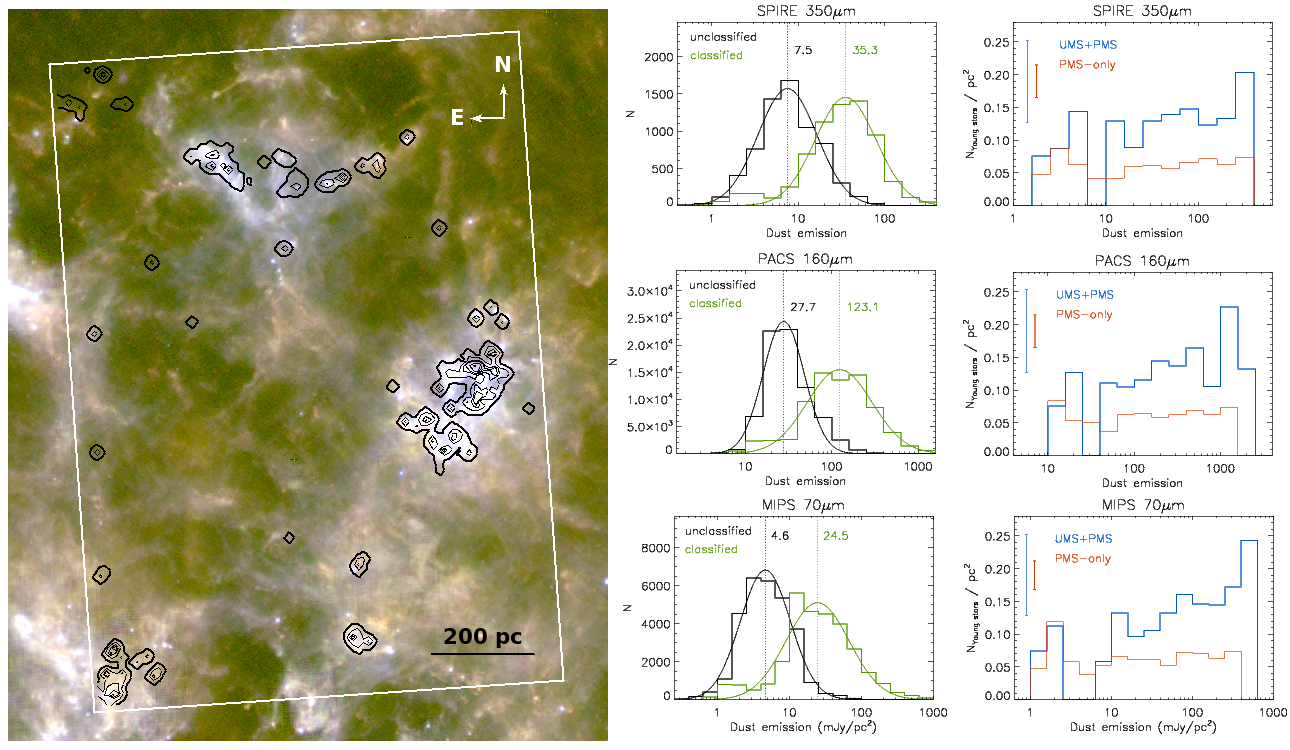}
   \caption{\textbf{Left panel:} Three-colour composite image with \textit{Spitzer} MIPS $70\,\mathrm{\mu m}$ in blue, \textit{Herschel} PACS $160\,\mathrm{\mu m}$ in green and \textit{Herschel} SPIRE $350\,\mathrm{\mu m}$ in red; the density contours are the same as those in Fig.\,\ref{fig:PMS_dist_class}. \textbf{Middle panels:} Dust emission distribution for image pixels in areas covered by the UMS+PMS and PMS-only classified elements, and by the same number of randomly selected unclassified elements. A Gaussian fit is plotted and the mean of the fit is indicated.  \textbf{Right panels:} Mean dust emission vs. young stars number density for the PMS and UMS+PMS classified elements. The error bars show the typical standard deviations within the dust emission bins.}
   \label{fig:PMS_dust}%
\end{figure*}
For $N_{\mathrm{PMS}} > 20$ the distribution can be approximated by a power law with a slope $\alpha(N_{\mathrm{PMS}}) = -0.86 \pm 0.12$. Studies of cluster mass distributions have shown that high mass clusters are less numerous than low mass ones \citep[e.g.][]{Zhang1999,Hunter2003,Grijs2008}, with a slope $\alpha(M)\,\sim\,-1$.
In a histogram using equal $\log{M}$ intervals, $\alpha(M) =-1$ is equivalent to a mass distribution function $n(M)\mathrm{d}m \propto M^{-\beta}\mathrm{d}m$ with $\beta = -2$, which is the slope expected for a scale-free hierarchical star formation scenario governed by turbulence \citep[e.g.][]{Fleck1996,Elmegreen2008}. Converting $N_{\mathrm{PMS}}$ obtained from our method into structure masses would yield very uncertain estimates due to the significant age dependence of the sensitivity limit of our VMC-based method (see Sec.\,\ref{subsec:synthetic_clusters}). For predominantly ``old'' PMS populations $\alpha(M) > \alpha(N_{\mathrm{PMS}})$ is expected, since a large fraction of PMS structures will have low PMS counts; for predominantly young PMS populations, we could expect $\alpha(M) < \alpha(N_{\mathrm{PMS}})$.
A more thorough discussion of the mass distribution should take the ages of the populations into account.

\subsection{Comparison with dust emission maps}
\label{subsec:dust_emission}

Since dust emission should be proportional to the product of dust mass and input stellar radiation (due to the energy balance between absorption and emission), one would expect a strong correlation between the number of young stars and dust emission. Bright far-infrared (FIR) emission is usually associated with high star formation rates (see \citealt{Casey2014} for a comprehensive overview); it originates from the young stars' radiation that is processed by dust from the remnants of their natal molecular clouds and re-emitted at longer wavelengths. Indeed, \cite{Skibba2012} reported that some regions with bright dust emission in the MCs coincide with known star forming regions. Moreover, in M33 \textit{only} young structures ($<100$\,Myr) were found to correlate with FIR surface brightness \citep{Javadi2017}. Given that the presence of PMS stars is an indicator of recent star formation we examine the relation with FIR emission in regions covered by the UMS+PMS and PMS-only classified elements. 

We make use of data from the SAGE and HERITAGE surveys in six FIR bands ranging from 70 to $500\,\mathrm{\mu m}$ \citep{Meixner2006, Meixner2013}. Figure \ref{fig:PMS_dust} (left) shows the PMS density contours overplotted onto an RGB image (70, 160 and $350\,\mathrm{\mu m}$) of the pilot field.  PMS-only and UMS+PMS classified elements are located along ridges and filamentary structures with bright dust emission. The only significant exceptions are Region A and some isolated PMS-only elements. As mentioned previously, the weak dust emission for Region A is in agreement with the inferred comparatively old age of the stellar populations.

    \begin{figure}
		\resizebox{\hsize}{!}{\includegraphics{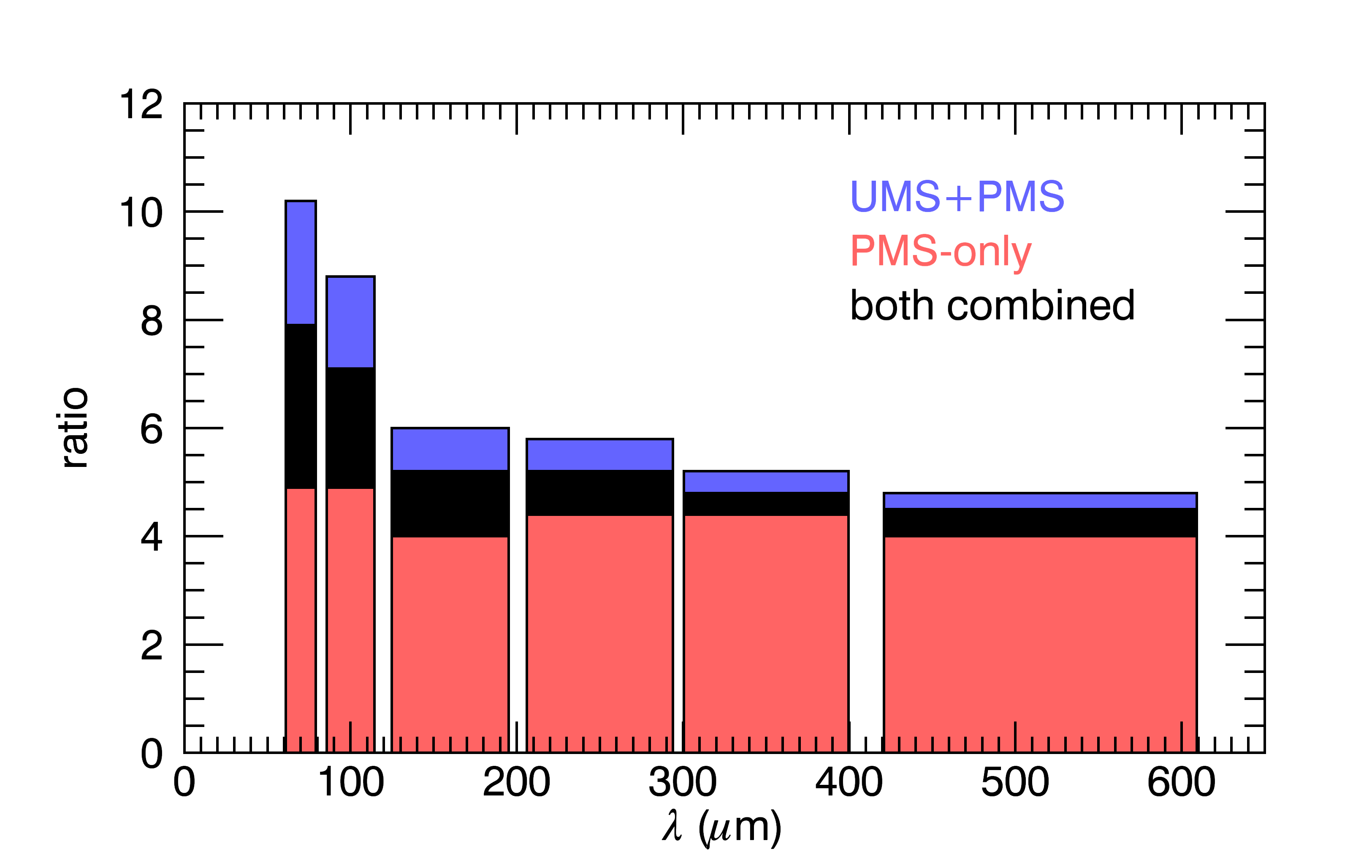}}
		\caption{Ratios of the mean dust emission observed 
        for classified elements and the field average. The widths of the bars mimic the filter bandwidths.
        }
		\label{fig:em_ratios}
	\end{figure}
The concentration of elements with PMS populations in regions with bright dust emission is explicitly shown in the middle histograms of Fig.\,\ref{fig:PMS_dust}. The dust emission distributions for grid elements associated with PMS populations and a randomly-selected sample of unclassified elements are clearly distinct; in particular, the strongest observed emission is always associated with young stellar populations. 

Figure \ref{fig:PMS_dust} (right) shows the dust emission versus the average number density of young stars for the same three FIR bands, separating PMS-only and UMS+PMS classified grid elements. We compute UMS densities for the UMS+PMS elements similarly to the PMS densities (Sec.\,\ref{subsec:PMS_count}) to obtain the total density of young stars.
We calculate the mean FIR emission for every classified element and group them in  0.2\,dex bins; for every bin the average number density of young stars is determined. For the UMS+PMS elements we observe a relatively weak but consistent trend towards higher stellar number densities with increasing dust emission for all wavelengths.  \cite{Hony2015} found a positive correlation between the number density of young stars -- UMS and PMS stars identified using HST photometry \citep{Gouliermis2006b} -- and dust surface density in the prominent star forming complex NGC\,346 in the SMC. Their analysis, on scales comparable to the size of our grid elements, is in agreement with the trend in Figure \ref{fig:PMS_dust} (right).
For PMS-only elements, however,  we do not observe an increase of stellar density with dust emission; less clustered low/intermediate mass PMS populations are likely more affected by the limitations of the VMC data.

Depending on the wavelength, the emission associated with the UMS+PMS and PMS-only classified elements is between four and eight times brighter than the field average (Fig.\,\ref{fig:em_ratios}); the strongest enhancement is found for the shorter wavelengths. Furthermore, the ratios for the PMS-only sample (red) are essentially independent of the wavelength,
while the ratio for the UMS+PMS sample (blue) increases significantly towards shorter wavelengths. This is evidence that the dust is heated primarily by the young massive UMS stars, while the PMS populations contribute very little. This is consistent with studies that found that warm dust follows the distribution of massive stars in the LMC \citep{Bernard2008}, as well as in Galactic star forming complexes \citep{Preibisch2012, Roccatagliata2013}. As a comparison, for the UMS-only classified elements (not shown) the dust emission is between 1.3 and 1.8 times higher than the pilot field average.

The analysis in this section confirms the important role that young UMS stars play in heating the dust.

\section{Summary and conclusions}
\label{sec:summary}

In this paper we presented a method to identify PMS populations ($\geqslant 1\,\mathrm{M_{\sun}}$) using data from the VMC survey.
The method applies Hess diagram analysis in the $K_s/(Y-K_s)$ space, including corrections for reddening and completeness, to disentangle young stellar populations from the underlying field; this analysis is performed independently on individual fixed-size spatial grid elements. 
Young populations are identified as density excesses (with respect to the field population) in pre-defined regions of the differential Hess diagrams. Depending on the location of these density excesses with respect to theoretical expectations (i.e. isochrones) we classify the population within a grid element into one of the four classes: PMS-only (population of young low-mass stars), UMS+PMS (young population with well-sampled IMF across the mass range), UMS-only (predominantly older population dominated by a prominent main-sequence population), and ``old'' (population that displays a significant contribution from evolved stars, in particular populating the red giant branch).

We apply our method to a $\sim 1.5\,\textrm{deg}^2$ VMC pilot field (LMC\,7\_5) and summarise our findings below:

\begin{itemize}

\item Tests with synthetic clusters explore the sensitivity of the method in the age range 1\,Myr -- 1\,Gyr and the mass range $250\,\mathrm{M_{\sun}}$ -- $3000\,\mathrm{M_{\sun}}$. We find that PMS populations can be identified up to an age of $\sim\,10$\,Myr for cluster masses $>1000\,\mathrm{M_{\sun}}$. Beyond 10\,Myr any remaining PMS populations are below the VMC sensitivity limit. The sensitivity increases towards younger ages: PMS populations with ages $\leqslant 2$\,Myr can be detected for clusters with masses down to $250\,\mathrm{M_{\sun}}$.

\item We \textit{detect} a total number of $2256\pm 54$ PMS stars in the pilot field. The most populous region is the N\,44 complex, which has $1000\pm 38$ PMS stars. This estimate must be taken as a lower limit, due to incompleteness in the VMC data and the sensitivity limit of our method.

\item The spatial distribution of elements with PMS populations is clearly inhomogeneous and clustered. UMS+PMS elements are almost exclusively found in large groups, while the PMS-only elements are more dispersed and often located in the outskirts of large star forming complexes. 

\item Large star forming complexes consist of multiple high stellar density peaks, with the highest densities found in the N\,44 complex. Overall, we detect 31 PMS structures which show a number distribution that can be approximated by a power-law with a slope of $-0.86\pm 0.12$.
A mass distribution with this slope would be broadly consistent with a hierarchical star formation scenario governed by turbulence.

\item The PMS populations are mostly located along ridges with intense dust emission in the FIR (70 -- $500\,\mathrm{\mu m}$). We observe a correlation between the dust emission and the number of young stars for the UMS+PMS elements. This is not the case for the PMS-only elements that lack UMS stars. In fact, dust emission is around four to eight times brighter for the UMS+PMS elements than in quiescent regions; at the shortest wavelengths the emission can be as much as 10 times brighter. This is likely due to dust heating by the radiation from the young UMS stars. 

\item Our analysis recovers all known star formation complexes in this field, and for the first time reveals their true spatial extent.

\item In the south-eastern corner of LMC\,7\_5 we discover a significant intermediate/low mass PMS population, that is likely associated with the wider N\,148 star forming complex. Comparison with synthetic cluster Hess CMDs suggests a very young age ($\sim 1$\,Myr) for this population. This population is co-spatial with significant CO emission.
\end{itemize}
 
Our method clearly shows the potential of the VMC survey to identify and characterise intermediate/low-mass young stellar populations on the scale of the whole Magellanic System. We are working on applying our method to other LMC and SMC VMC tiles.

\begin{acknowledgements}
This work is based on observations obtained with VISTA under ESO-program ID 179.B-2003. We thank the Cambridge Astronomy Survey Unit (CASU) and the Wide Field Astronomy Unit (WFAU) in Edinburgh for providing calibrated data products through the support of the Science and Technology Facility Council (STFC). This research has made use of the SIMBAD database, operated at CDS, Strasbourg, France, and of the NASA Astrophysics Data System (ADS) bibliographic services.  We thank Marta Sewi{\l}o for sharing her extensive compilation of LMC images and catalogues. Viktor Zivkov acknowledges studentships from ESO and the Faculty of Natural Sciences, Keele University, UK. We thank the anonymous referee for the constructive comments.
\end{acknowledgements}

\bibliographystyle{aa}
\bibliography{LMC1paper.bib}

\appendix

\section{Spatial distribution of the older population}
\label{sec:dist_old}

\begin{figure*}
   \centering
   \includegraphics[width=17cm]{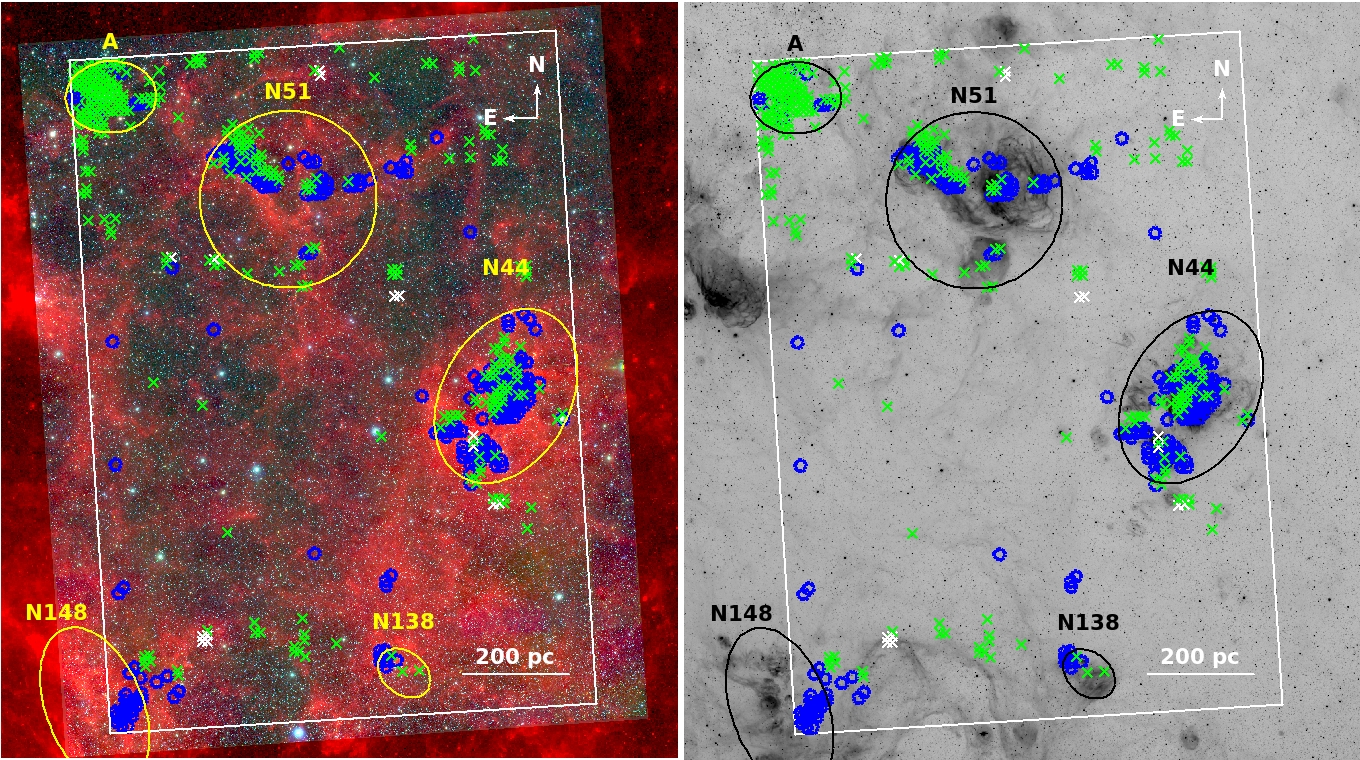}
   \caption{\textbf{Left:} Three-colour composite image with VMC $Y$ (blue) and $K_s$ bands (green), and \textit{Spitzer} IRAC $8.0\,\mathrm{\mu m}$ (see also Fig.\,\ref{fig:PMS_dist_class}). Green and white crosses mark the locations of elements classified as UMS and ``old'', respectively. The UMS+PMS and PMS-only classified elements are included as blue circles for ease of comparison. \textbf{Right:} As the left panel, but using an inverted grey-scale image of the $\mathrm{H\alpha}$ emission.}
   \label{fig:UMSold_dist_class}%
\end{figure*}
    \begin{figure}
		\resizebox{\hsize}{!}{\includegraphics{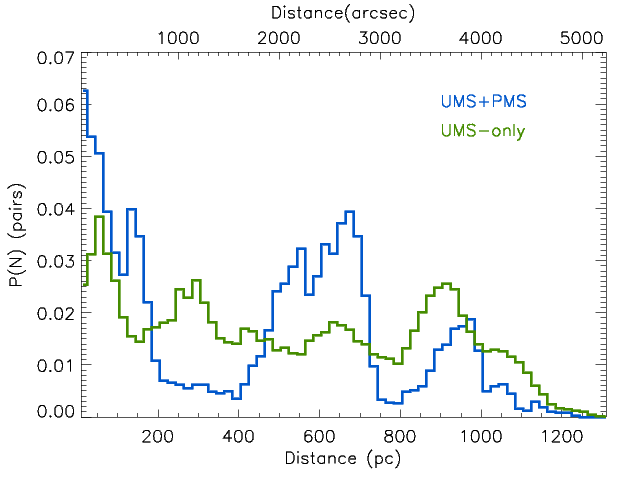}}
		\caption{Normalized distance distribution for all possible grid element pairs classified as UMS+PMS (blue) and UMS-only (green) (see also Fig\,\ref{fig:TPCF_grid}).} 
		\label{fig:TPCF_old}
	\end{figure}
In Sec.\,\ref{subsec:spa_distribution} the spatial distribution of elements showing a PMS signature was analysed. We found especially pronounced clustering for UMS+PMS classified grid elements, which represent young and relatively massive populations. Here we investigate the spatial distribution of the 298 grid elements classified as UMS-only. They contain comparatively old  (10\,$-$\,300\,Myr; see Sec.\,\ref{sec:strategy}) populations. 

Figure \ref{fig:UMSold_dist_class} shows the location of the grid elements classified as UMS-only and ``old''. Their spatial distribution is more scattered across the pilot field compared to the UMS+PMS elements. Consequently, more UMS-only elements are either isolated or located in small groups outside the known complexes. The concentration of UMS-only elements in region A is noticeable, further confirmation of the population's comparatively older age. The 14 grid elements classified as ``old'' are scattered across the field and are neither co-spatial with young grid elements nor with any significant dust or $\mathrm{H\alpha}$ emission. Twelve of them seem associated with known LMC clusters \citep{Bica2008}.

Fig.\,\ref{fig:TPCF_old} is equivalent to Fig.\,\ref{fig:TPCF_grid} (bottom), but shows the distance distributions for all possible pairs of UMS+PMS and UMS-only classified elements.
Some clustering is present for UMS-only elements, however it is much less pronounced than that for the UMS+PMS classified elements; UMS-only elements show a comparatively smooth distribution. This is in agreement with the temporal evolution of young stellar structures which are observed to disperse on long timescales \citep[e.g.][]{Sun2017b}.

\section{Reddening correction}
\label{sec:reddeningcorrection}
    \begin{figure}
		\resizebox{\hsize}{!}{\includegraphics{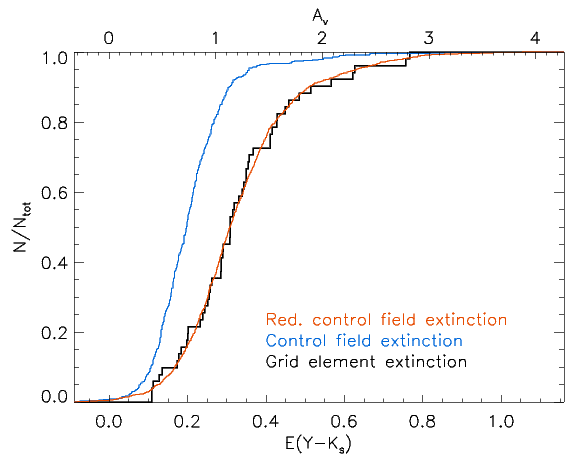}}
		\caption{Cumulative colour distribution of RC stars for the grid element under investigation (black line), and for the corresponding control field before and after the reddening correction (blue and red lines, respectively). RC stars are defined by the RC selection box in Fig.\,\ref{fig:hess_basic}. The intrinsic RC colour adopted is $(Y-K_s)_0 = 0.84$\,mag \citep{Tatton2013}.} 
		\label{fig:red_correction}
	\end{figure}
Before comparing a grid element with its corresponding control field, we redden the control field according to the distribution of RC stars in the grid element (see Sec.\,\ref{subsec:maps}). Figure\,\ref{fig:red_correction} shows how this correction is performed. The blue line represents the cumulative colour distribution of RC stars in the control field. Since the control field combines several grid elements, it contains many RC stars creating a very smooth distribution. The steep slope indicates a compact RC with not much differential reddening. The black line shows the cumulative distribution for the grid element being studied. The slope is less steep due to a more spread out RC. Especially the top $\sim 25\,$\% of the most strongly reddened stars suggests substantial differential reddening.

After applying the reddening correction to the control field population, its RC distribution (red line) closely follows the grid element's RC distribution. Differential reddening is accounted for, which would not have been possible by simply adopting a fixed reddening value for the control field stars.

\section{Images of selected star forming complexes}
\begin{figure*}
   \centering
   \includegraphics[width=17cm]{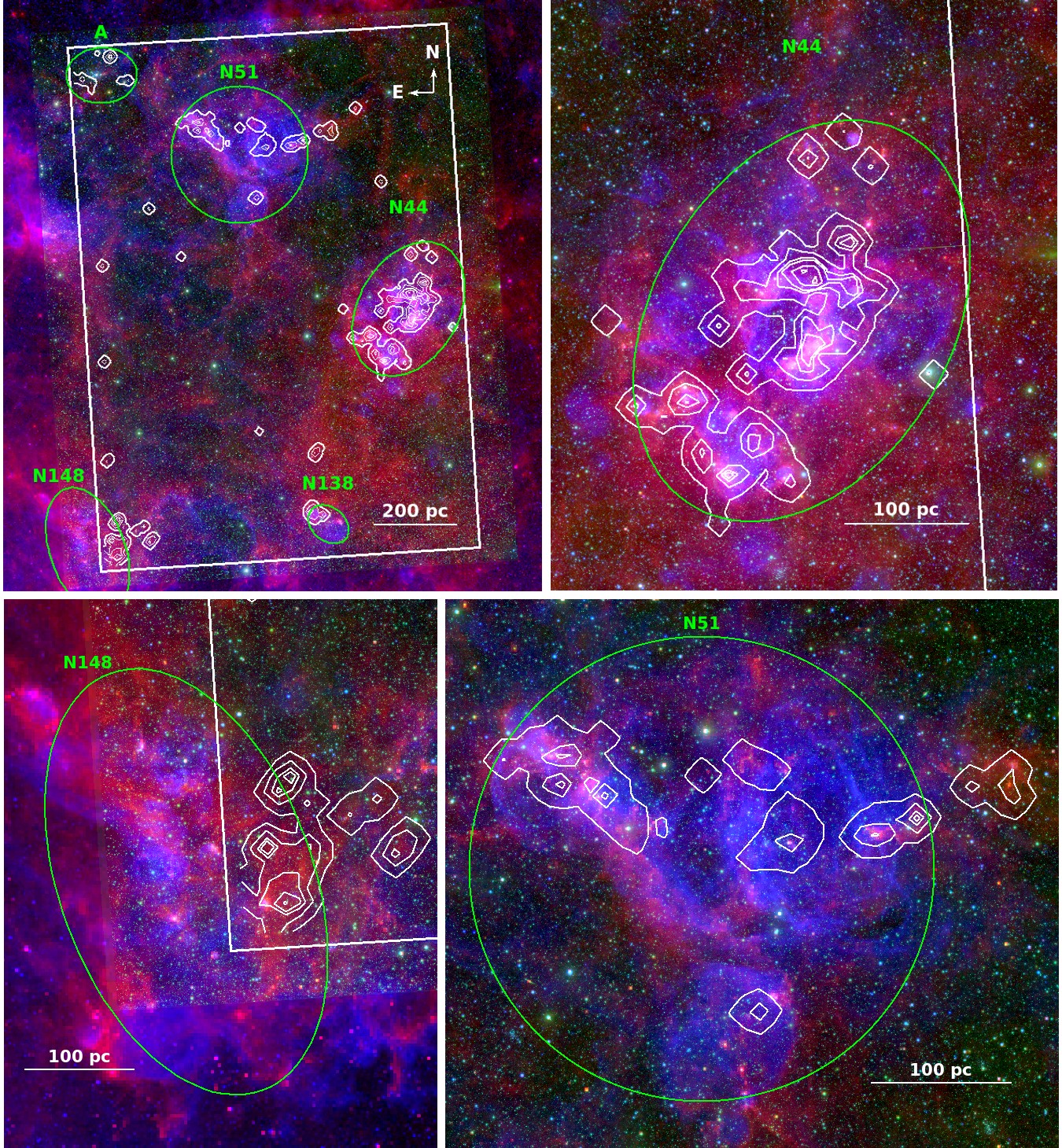}
   \caption{Three-colour composite image (top left) of the analysed area with $\mathrm{H\alpha}$ emission in blue, $K_s$ ($2.15\,\mathrm{\mu m}$) in green, and \textit{Spitzer} IRAC $8\,\mathrm{\mu m}$ in red. Regions and contours are the same as in Fig.\,\ref{fig:PMS_dist_class}. Other panels show enlarged views of selected star forming complexes: N\,44 (top right), N\,148 (bottom left) and N\,51 (bottom right).}
   \label{fig:zoom_complexes}%
\end{figure*}

Figure\,\ref{fig:zoom_complexes} presents close-up views of the three most populous regions in this LMC tile (see Tab.\,\ref{table:3}). It shows ionized gas ($\mathrm{H\alpha}$), stellar content ($K_s$-band), and hot dust ($8\,\mathrm{\mu m}$), together with the PMS density contours as calculated in Sec.\,\ref{subsec:PMS_count}.

\end{document}